%% file: paper2.tex
\documentclass[useAMS,usenatbib]{mnras}
\usepackage{amsmath}
\usepackage{aas_macros}
\usepackage{times}
\usepackage{amssymb}
\usepackage{placeins}
\usepackage{rotating}
\input{epsf}
\usepackage{color}

\bibpunct{(}{)}{;}{a}{}{,}

\def\imfs{$\alpha_3$}

\def\slug{{\scriptsize{\sc{SLUG}}}}
\def\msun{\rm{M}_\odot}
\def\clus{\texttt{cluster\_slug}}
\def\ha{{\rm H}$\alpha$}

\title[The IMF of star clusters with SLUG and LEGUS]
{Exploring the IMF of star clusters: a joint SLUG and LEGUS effort}
\author[G. Ashworth, M. Fumagalli, M. R. Krumholz, et. al.]
{G. Ashworth $^{1,2}$, M. Fumagalli $^{1,2}$, M. R. Krumholz $^3$, 
Angela Adamo $^4$, D. Calzetti $^5$,
\newauthor R. Chandar$^{6}$, M. Cignoni $^{7}, $D. Dale $^8$, B. G. Elmegreen $^9$, 
J. S. Gallagher III $^{10}$, 
\newauthor D. A. Gouliermis $^{11,12}$, K. Grasha $^{5}$, E. K. Grebel $^{13}$, K. E. Johnson $^{14}$, J. Lee$^{15}$, 
\newauthor M. Tosi $^{16}$, A. Wofford $^{17}$\\
$^1$Institute for Computational Cosmology, University of Durham, South Road,
Durham DH1 3LE, UK\\
$^2$Centre for Extragalactic Astronomy, University of Durham, South Road,
Durham DH1 3LE, UK\\
$^3$Research School of Astronomy \& Astrophysics, Australian National University, Canberra, ACT 2611, Australia\\
$^4$ Department of Astronomy, Stockholm University, Stockholm, Sweden\\ 
$^5$Department of Astronomy, University of Massachusetts-Amherst, Amherst, MA 01003, USA\\
$^{6}$Dept. of Physics and Astronomy, University of Toledo, Toledo, OH\\
$^7$Department of Physics, University of Pisa, Largo Pontecorvo, 3 Pisa, I-56127, Italy\\
$^8$Department of Physics \& Astronomy, University of Wyoming, Laramie, WY 82071, USA\\
$^9$IBM Research Division, T. J. Watson Research Center, Yorktown Heights, NY 10598, USA\\
$^{10}$Department of Astronomy, University of Wisconsin-Madison, Madison, WI 53706, USA\\
$^{11}$Zentrum f{\"u}r Astronomie der Universit{\"a}t Heidelberg, Institut f{\"u}r Theoretische Astrophysik, 69120 Heidelberg, Germany\\
$^{12}$Max Planck Institute for Astronomy, Heidelberg, Germany\\
$^{13}$Astronomisches Rechen-Institut, Zentrum f{\"u}r Astronomie der Universit{\"a}t Heidelberg, 69120 Heidelberg, Germany\\
$^{14}$Dept. of Astronomy, University of Virginia, 530 McCormick Road, Charlottesville, VA 22904, USA\\
$^{15}$Space Telescope Science Institute, Baltimore, MD, USA\\
$^{16}$INAF--Osservatorio Astronomico di Bologna, Bologna, Italy\\
$^{17}$Instituto de Astronom{\'i}a, UNAM, Ensenada, CP 22860, Baja California, Mexico\\}
\date{Submitted to MNRAS}
\pagerange{\pageref{firstpage}--\pageref{lastpage}} \pubyear{}

\date{Accepted XXX. Received YYY; in original form ZZZ}

\pubyear{2017}

\begin{document}

\topmargin = -0.5cm

\maketitle

\label{firstpage}

\begin{abstract}
  We present the implementation of a Bayesian formalism
  within the Stochastically Lighting Up Galaxies (\slug) stellar population
  synthesis code, which is designed to investigate variations in the initial mass function (IMF) of star clusters.
  By comparing observed cluster photometry to large libraries of clusters simulated with a continuously varying IMF,
  our formalism yields the posterior probability distribution function (PDF) of the cluster mass, age,
  and extinction, jointly with the parameters describing the IMF.  We apply this formalism to a sample of
  star clusters from the nearby galaxy NGC~628, for which broad-band photometry in five filters is available as part of the
  Legacy ExtraGalactic UV Survey (LEGUS). After allowing the upper-end slope of the IMF (\imfs) to vary, we recover PDFs
  for the mass, age, and extinction that are broadly consistent with what is found when assuming an invariant Kroupa IMF.
  However, the posterior PDF for \imfs\ is very broad due to a strong degeneracy with the cluster mass, 
  and it is found to be sensitive to the choice of priors, particularly on the cluster mass. We find only a modest
  improvement in the constraining power of \imfs\ when adding \ha\ photometry from the companion H$\alpha$-LEGUS survey.
  Conversely, H$\alpha$ photometry significantly improves the age determination, reducing the frequency of multi-modal PDFs.
  With the aid of mock clusters we quantify the degeneracy between physical parameters,
  showing how constraints on the cluster mass that are independent of photometry can be used to 
  pin down the IMF properties of star clusters. 
\end{abstract}

\begin{keywords}
  methods: statistical -- stars: luminosity function, mass function -- galaxies: star clusters: general --
  galaxies: individual: NGC628.
\end{keywords}

%==============================================
\section{Introduction} \label{sec:introduction}
%==============================================
A core parameter that is common to many astrophysical problems is the stellar initial mass function (IMF), which describes how stellar masses are distributed across a group of stars at the time of their birth. Constraining the form and evolution of the IMF as a function of galaxy properties is of great importance in our theoretical understanding of how star formation proceeds in galaxies and star clusters. Moreover, as the form of the IMF will dictate the makeup of a given star cluster, and hence its mass-to-light ratio, any inference an observer makes about the properties of a given unresolved stellar population via photometry will be strongly dependent on the assumed shape of the IMF \citep[see e.g.][]{Bell2003-Luminosity}.

The IMF is traditionally accepted to be universal, invariant both in time and with respect to the properties of the local environment. Two commonly adopted functional forms that describe the IMF are the broken power law of \cite{Kroupa2001-IMF} and the lognormal form of \cite{Chabrier2003-IMF} which features a power law tail at the high mass end ($ M \ge 1\,\msun $). Both these functional forms possess a high-mass slope that is consistent with that of the power-law IMF originally  presented by \cite{Salpeter1955-IMF}, with index $\alpha=-2.35$, . This form of the IMF appears to hold in most nearby environments observed to date \citep{Massey1995-IMF,Bastian2010-IMF}.

Although theoretical predictions exist for the variation of the IMF with, for example, metallicity and gas temperature \citep{Adams1996-IMFTheory,Bonnell2006-IMFJeansMass,Krumholz2011-FragIMF}, at present there is no widely accepted evidence of IMF variation.
There is, however, an increasing body of literature supporting the notion of a non-universal IMF, with variations across different environments. For instance, \cite{vanDokkum2010-Elipticals} found evidence suggesting a bottom-heavy IMF in early-type galaxies based on the strength of distinctive absorption lines (NaI doublet and Wing-Ford molecular FeH band), which can be modelled with an overabundance of low mass stars in massive galaxies. Similar detections have been reported in the more recent literature \citep{Conroy2012-IMFinEarlyTypes}, with further evidence in support of a steeper IMF in massive elliptical galaxies coming from dynamical measurements \citep{Cappellari2012-IMFinEarlyTypes}. While both methods
consistently imply more bottom-heavy IMFs in massive galaxies, they appear to
disagree in detail when applied to individual galaxies rather than to a galaxy
population \citep[see][]{Smith2014-IMFVar}. However, a more recent study by \cite{Lyubenova2016-CALIFA} using consistent data sets  has not found this disagreement to be apparent.

At the other end of the galaxy mass function, recent observations have also revealed apparent correlations between the \ha-to-FUV luminosity ratio and the star formation rate (SFR) \citep{Lee2009-HaFUV,Lee2016-HaFUV} in dwarf galaxies, which can be attributed to variation in the high-mass end of the IMF. Similar suggestions of variation in the IMF have been made based on comparisons between the \ha-to-FUV luminosity ratio and \ha\ and R-band surface brightness in starburst galaxies \citep{Meurer2009-IMF}. Other indications of a variable IMF have been gleaned using techniques involving the \ha\ equivalent width and galaxy colours \citep{Gunawardhana2011-GAMA}, and chemical evolution arguments \citep{Matteucci1994-ChemEv1,Thomas1999-ChemEv1}. 

Attempts to explain these observations include the framework of the Integrated Galactic Initial Mass Function (IGIMF) \citep{Kroupa2001-IMF,Kroupa2003-IGIMF,Weidner2011-IGIMF,Kroupa2013-StellarSubStellar}. In this theory, the most massive stars allowed to form in a cluster (corresponding to the high-mass cutoff of the IMF) are correlated with the cluster mass, which in turn depends on the galaxy SFR. These correlations suppress the formation of massive stars in dwarf galaxies,
resulting in a lower \ha-to-FUV ratio, in line with the aforementioned observations.

However, the very same observations can be explained by alternative models that do not resort to deterministic variations of
the IMF. For example, stochasticity has been shown to suitably describe the \ha-to-FUV ratio correlation in dwarf galaxies without the need for an explicit correlation between star and cluster masses. Indeed, a combination of the random sampling of
clusters from a cluster mass function combined with a random sampling of stars from the IMF  \citep{Fumagalli2011-SLUGLetter},
appears to satisfactorily reproduce the observed \ha-to-FUV ratio in dwarf galaxies. Similar results have been reported by \citet{Calzetti2010-IMF} and \citet{Andrews2013-IMFinNGC4214}, who have studied the ratio of \ha\ to cluster mass, finding that the IMF upper end slope has no clear mass dependence \citep[see also][]{corbelli09}. \citet{Andrews2014-IMFinM83} find that stochasticity is similarly effective in explaining observations of the spiral galaxy M83. Additional challenges to the IGIMF theory are discussed, for instance, in \citet{Oey2013-SoloOBStars} and \citet{Lamb2016-RIOTS}. These authors present observations of what are likely to be solitary O and B stars with no visible evidence of escape from nearby clusters. Such an occurrence would violate the requirement of the IGIMF that the maximum stellar mass correlates with cluster mass.  However, previous indications of the possibilitiy of solitary OB stars \citep{deWit2004-SolitaryOB,deWit2005-SolitaryOB,Lamb2010-SolitaryOB} have been disputed by \citet{Gvaramadze2012-Runaways}, and more recently \citet{Stephens2017-OBStars} have made arguments in opposition to stochastic IMF sampling as a rebuttal of the IGIMF through observations of compact clusters of pre-main-sequence stars around massive stars previously thought to be isolated.

 This ambiguity surrounding the universality of the IMF calls for the development of new techniques and methods that explore the properties of the IMF in different environments. To this end, Bayesian techniques have recently been applied to stellar population synthesis (SPS) models, in an attempt to infer the parameters that describe the IMF. A particularly appealing feature of these Bayesian approaches is the possibility to marginalise over other ``nuisance'' parameters, and to explore to what extent the IMF is degenerate with respect to other parameters used in SPS models.
 
For example, \cite{Weisz2012-PHAT,Weisz2015-M31IMF} use Bayesian techniques combined with color-magnitude diagrams of resolved star clusters to derive a constant IMF high-end slope that is both steeper than the canonical Kroupa value of \imfs=-2.3 and constant in Andromeda.  
 \citet{Dries2016-BayesianIMF} have also recently presented an example of how Bayesian methods can be applied
 to this problem, successfully recovering the input parameters describing
 the IMFs in a selection of mock simple stellar populations produced with varying IMFs, provided
 however that a representative set of stellar templates is available during the analysis.

 Following a similar idea, in this work we aim to combine Bayesian techniques with the stochastic SPS code \slug\
 \citep[Stochastically Lighting Up Galaxies;][]{daSilva2012-SLUGi,daSilva2014-SLUGii,Krumholz2015-SLUGiii} to explore the
 IMF properties of star clusters.
 This analysis builds on the method described in \citet{Krumholz2015-SLUGandLEGUS}, who combined
 \slug\ with Bayesian inference to constrain the mass, age, and extinction of clusters from broad-band photometry. 
 In this work, expanding on their previous analysis, we  develop a formalism to also
 investigate whether the high-mass end of the IMF can be constrained through
 photometry of individual clusters, while tracking at the same time degeneracies between the other
 parameters that regulate the mass-to-light ratio of the stellar populations.
 To achieve this goal, we first need to extend the current capabilities of \slug, as we
 describe briefly in Section~\ref{sec:introvariable}, and in Appendix~\ref{sec:variable}. Specifically, we enable a new mode in \slug\
 through which libraries of simulated clusters can be constructed using a continuous distribution
 of arbitrary IMFs. These libraries can then be compared to actual observations using Bayesian
 techniques to infer the posterior probability distribution function (PDF) for the parameters that describe
 the cluster physical properties, including the IMF, within SPS models.

 Following the implementation of this extension in the \slug\ code, in Section~\ref{sec:legus},
 we apply this extended Bayesian formalism to photometric data on star clusters from the Legacy ExtraGalactic
 UV Survey (LEGUS). LEGUS is a Cycle 21 \textit{Hubble Space Telescope} Treasury program  \citep{Calzetti2015-LEGUSi}, designed to investigate star formation and its relation with the galactic environment across a representative sample of nearby galaxies within 12\,Mpc, on scales ranging from individual stars up to kiloparsec-scale structures. In addition to this application,
 in Section~\ref{sec:predict} we construct a library of mock star clusters, which we use to explore from a theoretical point
 of view how accurately the upper-end slope of the IMF (hereafter the $\alpha_3$ parameter) can be recovered using photometry from
 star clusters. In doing this, we also quantify the degeneracy between IMF parameters and other physical quantities, such as mass and age. Summary and conclusions follow in Section \ref{sec:conclusions}, where we discuss future improvements to our method and describe
 future experiments that could yield tight constraints on the IMF of star clusters. 
 Throughout this work, unless otherwise specified, magnitudes are in the Vega system. 

%==============================================================================

\section{The SLUG software suite} \label{sec:introvariable}

The core of the \slug\ suite is an SPS code that, unlike the majority of
traditional SPS codes, such as {\scriptsize{\sc{STARBURST99}}} \citep{Leitherer1999-STARBURST99}, includes the effects of stochastic sampling. Stochastic sampling of the IMF was pioneered in the {\scriptsize{\sc{MASSCLEAN}}} software of \cite{Popescu2009-MASSCLEAN1}. In \slug\, this stochasticity is achieved through generating
galaxies by drawing stars and clusters from underlying PDFs which define the IMF and the cluster mass function
(CMF) in the simulated galaxy. These populations are then evolved through time to the desired age, accounting for cluster disruption. We refer the reader to previous work for a detailed description of the algorithms
adopted in \slug\ \citep[e.g.][]{daSilva2012-SLUGi,Krumholz2015-SLUGiii}.
Briefly, in its basic ``cluster mode'', \slug\ constructs star clusters by drawing stars from
an IMF until a desired target mass is reached. This sampling method allows \slug\ to accurately
capture the effects of  stochasticity, which are particularly relevant for low mass clusters
\citep[e.g.][]{Cervino2002-Sampling,Elmegreen2002-StarFormation,Haas2010-IMFVar}.
Moreover, the ability to populate star clusters (or galaxies) by randomly drawing from the
IMF and CMF provides a convenient way to generate large
libraries of realistic clusters which are representative of the observed population in nearby
galaxies \citep[e.g.][]{Krumholz2015-SLUGiii}.
 
Along with the core SPS code, the \slug\ distribution includes the \clus\ software, which is a package designed for Bayesian
analysis of observed cluster photometry \citep{Krumholz2015-SLUGiii}.
Specifically, \clus\ uses Bayesian inference to recover the posterior PDFs
for physical parameters (e.g. mass, age, extinction) that influence the photometric
properties of a given star cluster. To this end, the \clus\ package makes use of a large
`training set' of \slug\ models that, through Bayesian analysis combined with kernel density
estimation (KDE) techniques \citep{daSilva2014-SLUGii,Krumholz2015-SLUGiii}, yields the PDFs
for the parameters of interest given the observed broad-band photometry of a stellar cluster. 

In this work, we extend this technique to the case of varying IMFs. Within the Bayesian framework of \clus, the posterior
PDFs of parameters describing the functional form of the IMF (for instance, the high end slope \imfs\ in the Kroupa IMF) can be
derived from cluster photometry, provided that \slug\ libraries containing a continuous distribution of clusters with respect
to the parameter of interest (e.g. \imfs) are available. To this end we have implemented a variable IMF capability in \slug, with which libraries of simulated clusters can be generated with continuously variable IMF parameters. This modification is discussed in detail in Appendix~\ref{sec:variable}.

%==============================================================================

\section{Applying SLUG to observed star clusters}\label{sec:legus}
Having enabled a variable IMF capability in \slug, we now combine a large library of simulated star clusters
with LEGUS broad-band photometry. This photometry is taken from a star cluster catalogue of the galaxy NGC~628. We use this combination to investigate whether it is possible to constrain the upper-end slope of the IMF with our Bayesian technique.

\subsection{The NGC~628 photometric catalogues}\label{sec:catalogue}

For this analysis, we make use of the multiwavelength imaging available from the LEGUS project.
As part of the LEGUS survey \citep[see][]{Calzetti2015-LEGUSi},
local volume galaxies have been imaged with five broad-band filters
using the Wide-Field Camera 3 (WFC3), with parallel observations from the Advanced Camera for Surveys (ACS). Of the fifty galaxies in the LEGUS sample, we select NGC~628 for our analysis. The filters used for NGC 628 are as follows: WFC3 F275W, F336W, and F555W, along with ACS F435W and F814W. These may be generally thought of as the NUV, U, B, V, and I bands.
In our study, beside broad-band photometry, we also make use of narrow-band H$\alpha$ data,
which have been obtained in the WFC3~UVIS~F657N filter as part of the \ha-LEGUS program (Chandar et al. in preparation).

NGC~628 is a nearby face-on grand design spiral galaxy of type SA(s)c at a distance of 9.9\,Mpc \citep{Calzetti2015-LEGUSi}.
Currently the most well studied LEGUS galaxy, it is one of the two galaxies discussed
in \cite{Krumholz2015-SLUGandLEGUS}. Thus, the properties of star clusters in NGC~628 have already been
analysed using the \slug\ suite for a constant IMF, which provides us with a reference point to which we can
compare our results when variations of the IMF are included. 
Two pointings are available for this galaxy (NGC~628 East and NGC~628 Centre), and here
we focus on the the East pointing (hereafter NGC~628E) as done in \citet{Krumholz2015-SLUGandLEGUS}.
Moreover, H$\alpha$ data are available for the NGC~628E pointing, which allows us to explore the effects
of including photometry that is sensitive to massive stars in our analysis.

We refer the readers to relevant LEGUS papers for a detailed description of
the data processing techniques, and particularly on how cluster catalogues are generated
\citep[e.g.][Adamo et al. 2017, submitted]{Calzetti2015-LEGUSi,Grasha2015}. Here we provide only
a brief description of the most relevant procedures.
We make use of clusters that have been detected in at least 4 filters. 
For all clusters, both an aperture correction
and correction for foreground Galactic extinction have been applied following standard procedures.
All the clusters have been visually classified, according to the LEGUS definition of class 1-3.
Class 1 objects are compact and centrally concentrated clusters,
class 2 objects are clusters with elongated surface-brightness profiles, and class 3 objects loosely comprise associations
or systems with multiple resolved peaks in surface brightness. We omit class 4 objects from our analysis, as they represent
for most part isolated stars, background galaxies, or artefacts.

With this selection, we are left with a cluster catalogue comprised of 259 clusters from NGC~628E, as in \cite{Krumholz2015-SLUGandLEGUS} and Adamo et. al. (2017, submitted). Of these, we further select 241 clusters with a detection in the WFC3~UVIS~F657N filter.
Of the clusters with positive detections in this filter, 16 are discarded from the following analysis as
visual inspection shows that their morphology is indicative of emission not arising from HII regions (Lee et. al. in preparation). 
Altogether, our selection results in a final catalogue containing 225 clusters in the LEGUS class 1-3.

\subsection{Building a library of $2\times 10^{8}$ simulated clusters}\label{sec:libbuild}

The first step towards applying our formalism to the LEGUS cluster catalogue is the
generation of a library that provides the training set required for Bayesian analysis in \clus.
Having enabled the variable IMF mode in \slug, we now trivially generate an extended library of simulated
star clusters, including a continuous range of IMFs with a varying upper-end slope ($\alpha_3$).
Our focus on the massive end of the IMF is justified by the availability of NUV and \ha\ photometry
in LEGUS and \ha-LEGUS. 

\input{table_reflib.tex}

To construct this reference library with a variable IMF, which we use in all subsequent calculations unless stated otherwise,
we extend the fiducial library described in \cite{Krumholz2015-SLUGandLEGUS}\footnote{The library is publicly available
  at www.slugsps.com and contains simulated cluster photometry in the selection of filters used in the LEGUS observations
  of NGC~628}
to include a new dimension for the variable upper-end slope $\alpha_3$.
This library, which we dub {\tt pad\_020\_vkroupa\_MW}, contains $2\times 10^8$ clusters distributed in mass
as
\begin{equation}
p_M (\log M) \propto
\begin{cases}
1, & 2 < \log M < 4 \\
10^{-(\log M - 4)}, & 4 \le \log M < 8 \:,
\end{cases}
\label{eq:sampledist1}
\end{equation}
where the mass $M$ is in units of $\msun$.
As in \cite{Krumholz2015-SLUGandLEGUS}, this choice is purely dictated by computational efficiency, as low
mass clusters are cheaper to simulate. Furthermore, with this weighting, we generate the largest number
of clusters in the mass range where stochasticity is dominant so as to appropriately sample the dispersion
in the population arising from random sampling. As described below, however,
the choice of weighting does not affect our analysis as we can apply appropriate weighting schemes to
produce arbitrary mass distributions.
Similarly, we assume an age distribution defined by
\begin{equation}
p_T (\log T) \propto
\begin{cases}
1, & 5 < \log T < 8 \\
10^{-(\log T - 8)}, & 8 \le \log T < \log T_{\mathrm{max}}\:,
\end{cases}
\label{eq:sampledist2}
\end{equation}
which results in the generation of more young clusters, in which stochastic effects associated with massive stars are more prominent.
In these equations the time $T$ is in units of years,
and $T_{\mathrm{max}}$ is the maximum age for which stellar tracks are available. 

The clusters are populated with stars drawn from a variable IMF.
Here, we assume a Kroupa-like IMF, defined as a broken power-law comprised of three segments:
\begin{equation}
\alpha_i = 
\begin{cases}
-0.3, & 0.01 \leq M < 0.08,~  i=1 \\
-1.3, & 0.08 \leq M < 0.5,~ i=2 \\
-3.0 \leq \alpha_3 \leq -1.5, & 0.5 \leq M < 120,~ i=3 \:,
\end{cases}
\label{eq:imfslopes}
\end{equation}
where $\alpha_i$ is the power law index of the $i^{\rm{th}}$ segment and the stellar masses are in units of $\msun$. The value of the
third segment slope (\imfs) is drawn from a flat distribution between $-3.0$ and $-1.5$ at the start of each realisation.

Photometry is computed using Padova tracks including thermally pulsating AGB stars \citep{Vassiladis1993-PadovaAGB,Girardi2000-Padova,Vazquez2005-Padova}, which sets the
maximum age $T_{\mathrm{max}}= 15~$ Gyr. We assume a metallicity $Z=0.02$, consistent with the solar value, and 
we apply a uniform extinction to each cluster assuming a Milky Way extinction curve \citep{Cardelli1989-MWExt}, which we normalise
by drawing the visual extinction $A_{\rm V}$ from a flat distribution in the range $0 < A_{\rm V} < 3$. 
In this work we use the default {\sc SB99} spectral synthesis mode of \slug, which emulates the behaviour of {\scriptsize{\sc{STARBURST99}}} in choosing which atmosphere model is used for a given star.
  Details of this implementation can be found in Appendix A2 of \citet{Krumholz2015-SLUGiii}.
  While this set of atmospheres is commonly adopted in SPS modelling, we note that our results
  will naturally depend on this choice.

We further compute the nebular contribution to the broad-band photometry, assuming that
$50\%$ of ionising photons are converted into nebular emission (see also Section \ref{sec:halegus}). 
The choice of these parameters is dictated by the properties of NGC~628 (see also Section~\ref{sec:haprep}).
\citet{Krumholz2015-SLUGandLEGUS} demonstrate that the results of \clus\ analysis are mostly insensitive to the choice of tracks, extinction curve, and metallicity, and so for this study we make use of only one choice for each of these parameters. Metallicity, choice of tracks, and ionisation efficiency do become important when considering clusters with ages less than $3\,\rm{Myr}$ however.

We do not include a treatment of binary stars or stellar rotation in this study. The presence of binaries (especially among massive stars) has been shown to produce appreciable effects on the integrated light of star clusters \citep{Eldridge2009-Binaries}, resulting in a less luminous and bluer stellar population \citep{Li2008-Binaries}. However, as their inclusion requires several additional parameters,
  binaries are currently not treated in \slug. Rotation, which likewise affects results of
SPS models \citep{Vasquez2007-Rotation,Levesque2012-Rotation,Leitherer2014-Rotation}, can be included through the use of the latest Geneva tracks \citep{genevarot} available in \slug.
However, we choose to consider the non-rotating Padova tracks to include the AGB treatment currently not present for the Geneva tracks with rotation. The effects of rotation can be as much as a $50\%$ increase in luminosity of O stars, along with a factor of 2 increase in luminosity between the UV and NIR \citep{Vasquez2007-Rotation,Levesque2012-Rotation}. However, the effects are only significant if all the stars have initial rotation velocities of $\approx 300\,{\rm km}\,{\rm s}^{-1}$ \citep{Eldridge2009-Binaries}.

Throughout this analysis, we also make use of the fiducial library of \cite{Krumholz2015-SLUGandLEGUS} with
a constant Kroupa IMF to compare our results to the case of a non-variable IMF. 
A summary of the properties of these reference libraries is given in Table~\ref{tab:simparam}.
Our library contains $2\times 10^8$ clusters, which is a factor of $20$ larger than the
  fiducial library of \cite{Krumholz2015-SLUGandLEGUS}. This choice is dictated by the need to
  increase the number of clusters to account for the extra free parameter of the IMF slope \imfs,
  while keeping the library size commensurate to the available computational resources.
  We have performed tests using smaller libraries of $10^8$ clusters, finding that
  we have reached a satisfactory convergence.

\subsection{Bayesian techniques and choice of priors}\label{sec:bayprio}

With a library of $2\times 10^8$ clusters featuring variable IMFs in hand, we next make use of the Bayesian analysis tool \clus\
to handle the computation of the posterior PDFs of the $\alpha_3$ parameter.
A description of the algorithms adopted in \clus\ has been presented in \citet{daSilva2014-SLUGii} and
\citet{Krumholz2015-SLUGiii}, and we refer the readers to this work for details.
Briefly, the Bayesian analysis tool \clus\ takes a set of absolute magnitudes $M_F$ over a given set of filters, along with the
associated errors $\Delta M_F$. By means of a large library of \slug\ models,
\clus\ exploits kernel density estimation to map the observed photometry into a 
PDF for the cluster physical parameters.
In previous work, the only parameters under consideration were the cluster mass $M$, age $T$, extinction $A_{\rm V}$.
Here, we further extend \clus\ to also output PDFs for IMF parameters that are chosen to be variable, which in our case is the \imfs\ parameter.
During this calculation we set the kernel density estimation bandwidth to $h=0.1\,\rm{dex}$ for the physical variables and to
$h=0.1\,\rm{mag}$ for the photometric variables, as in \citet{Krumholz2015-SLUGandLEGUS}. During testing, a shift to a smaller bandwidth of $h=0.05$ did not appreciably affect the results of the analysis.

As discussed in the previous section, the simulated clusters in our library are not weighted evenly in parameter space
for reasons of computational efficiency and to better sample the range of parameters where stochasticity is most relevant.
It therefore becomes necessary to ``flatten'' the input library, prior to feeding it into \clus.
This step is accomplished simply by inverting the weights chosen during the calculation of the library of \slug\ clusters.
After the library has been flattened, we further need to make a choice of the priors to be used during the Bayesian
analysis. Throughout this work, we explore the effects of two different choices of prior. Our first choice is a flat
prior, with $p_{\mathrm{prior}} = 1$. Our second choice is a prior in the form of: 
\begin{equation}
p_{\mathrm{prior}}(\textbf{x})\propto M^{-1}T^{-0.5},
\label{eq:k15prior}
\end{equation}
for $T > 10^{6.5}\,\mathrm{yr}$.
This functional form is physically motivated by the cluster mass distribution, which is observed to follow a
power-law with index $-2$. The time dependence accounts instead for the cluster disruption
\citep[see][]{Krumholz2015-SLUGandLEGUS}.  For young ages of $T \le 10^{6.5}\,\mathrm{yr}$, where cluster disruption
is largely ineffective, we switch to a prior $p_{\mathrm{prior}}(\textbf{x})\propto M^{-1}$. 
As evident from the above equations, we elect to adopt a flat prior for both the IMF high-end slope \imfs\ and the extinction $A_{\rm V}$ in both cases.

\begin{figure}
\centerline{\includegraphics[scale=0.42]{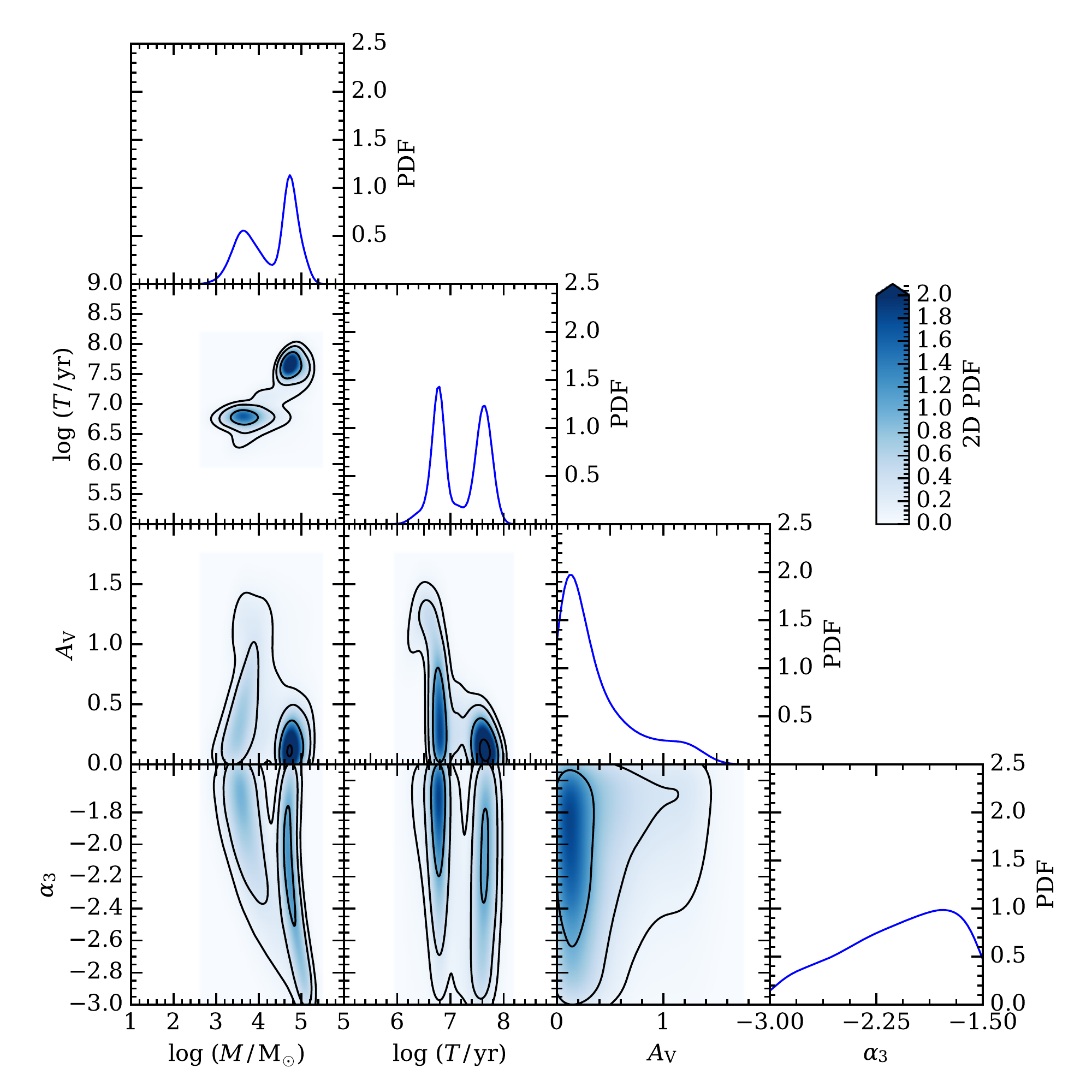}}
\centering
\caption{Example of a triangle plot for cluster ID~56 in NGC~628E, computed assuming the physically motivated priors from
  \protect\cite{Krumholz2015-SLUGandLEGUS}. The 1D posterior PDFs for $\log M$, $\log T$, $A_V$, and \imfs\
  that are constructed marginalising over all the other parameters are shown in the top panels of each column.
  The contour plots show instead the joint posterior PDFs, where the intensity reflects
  the probability density as indicated by the colour bar. In each panel, contours are spaced in
  intervals of 0.2 unit element. All the PDFs are normalised to have a unit integral.
}
\label{fig:cl56nhak15}
\end{figure}

\begin{figure}
\centerline{\includegraphics[scale=0.42]{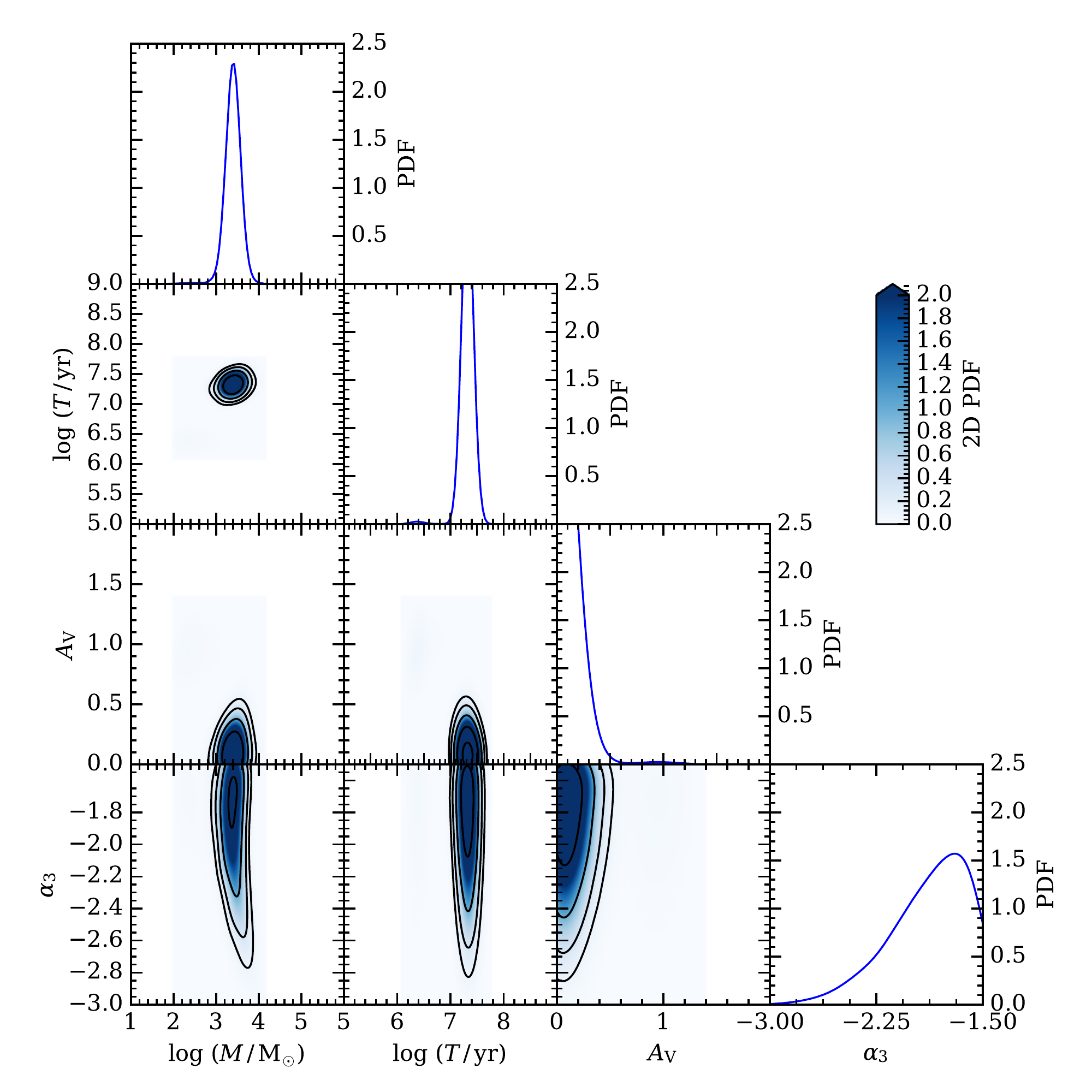}}
\centering
\caption{As Figure \ref{fig:cl56nhak15}, but for cluster ID~591.
}
\label{fig:cl591nhak15}
\end{figure}

\begin{figure}
\centerline{\includegraphics[scale=0.45]{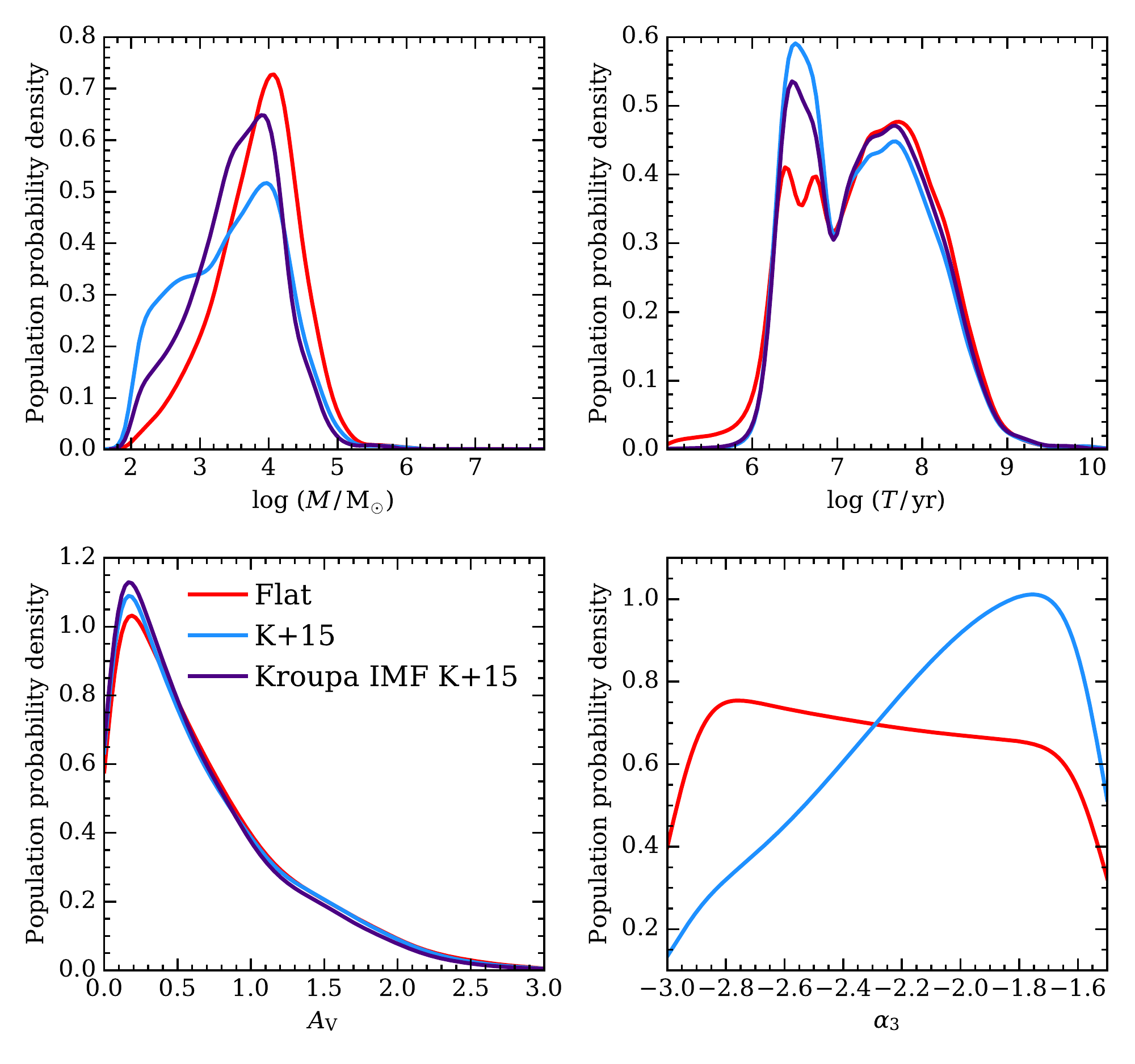}}
\centering
\caption{The cumulative posterior PDFs for the star clusters in our sample from  NGC~628E, computed assuming both flat priors (in red, labelled as Flat) and the physically motivated priors of \protect\cite{Krumholz2015-SLUGandLEGUS} (in blue, labelled as K+15). Also included are the cumulative PDFs calculated using the fiducial library of \protect\cite{Krumholz2015-SLUGandLEGUS} with a constant Kroupa IMF and physically motivated priors (in purple, labelled as Kroupa IMF K+15). The distributions are comparable overall. There is significant deviation between the two priors in the case of \imfs, with the \protect\cite{Krumholz2015-SLUGandLEGUS} priors having a significant preference towards more shallow IMF slopes.}
\label{cumulatives}
\end{figure}

\subsection{Analysis of NGC~628E with broad-band photometry}\label{sec:legusanalysis}

Our next step after constructing a library of simulated clusters with variable IMF is to apply our \clus\ analysis to the LEGUS broad-band photometry of NGC~628E. With our extension of the \slug\ code that now handles the case of
a variable IMF, we use the \clus\ package to calculate the one dimensional (1D) posterior PDFs of all the physical parameters of interest
(mass, age, extinction, and \imfs) for  the 225 clusters we have selected from NGC~628E. Two dimensional PDFs are
also computed, which are useful to explore correlations among parameters. 
At this stage, we apply both flat priors and the physically motivated priors previously presented in \citet{Krumholz2015-SLUGandLEGUS} (see Equation~\ref{eq:k15prior}). In this section, we begin our analysis by considering the results obtained
with the physically motivated priors, but in Section \ref{sec:priors} we explicitly investigate the effects
of the choice of priors. 

Before considering the cluster population as a whole, it is instructive to inspect the corner plots for two
example clusters, which we choose to bracket the range of behaviour seen in our analysis.
As a first example, we show the corner plot for cluster ID 56 in Figure~\ref{fig:cl56nhak15},
where the posterior PDFs are obtained using the physically motivated priors. This is the same cluster that is presented
as an example in Figure~8 of \cite{Krumholz2015-SLUGandLEGUS}, and having it as an example here allows us to make a direct comparison with
previous work. However, we note that it is not part of the reduced 225 cluster catalogue as it has no observed \ha\ emission. The shape of the posterior PDFs for mass, age, and extinction computed for a variable IMF slope
are quite similar to those recovered by \cite{Krumholz2015-SLUGandLEGUS} with a fixed IMF, with the exception of a
small excess of probability towards lower masses and ages. 
This cluster also happens to be representative as to the behaviour of a significant fraction of the clusters in our sample,
which exhibit multiple peaks in the posterior PDF for mass and age. Furthermore, and most significantly,
the 1D posterior PDF for \imfs\ is very broad, spanning the full range of possible values with a
modest preference for shallow slopes. An inspection of the joint posterior PDFs further reveals that
the IMF slope parameter is largely insensitive to age and extinction, and is degenerate with respect to mass.

At the opposite end of the spectrum, we show in Figure~\ref{fig:cl591nhak15} the case of a cluster
(ID~591) for which mass, age, and extinction are well constrained, as is clear from the
sharp posterior PDFs of these quantities. Despite the fact that these physical parameters are well
constrained, the posterior PDF of \imfs\ is still quite broad, with most of the probability
being contained in the interval $\alpha \approx (-2.3,-1.5)$. As was seen in cluster~56, the corner plot reveals a modest preference of shallow IMF exponents, with $\alpha \gtrsim -2.2$.

The features highlighted in these examples appear to be quite general of the entire sample. This is shown in
Figure \ref{cumulatives}, where we display the cumulative posterior PDFs for the entire sample, which we obtain by 
combining the 1D posterior PDFs for each cluster. It must be noted that we do not analyse the clusters jointly for this, and that the PDFs for the clusters are combined following our analysis. 

In addition to the two choices of prior, which we will discuss
below, we also include in this figure the cumulative PDFs calculated using the fiducial library of
\cite{Krumholz2015-SLUGandLEGUS}, assuming a constant IMF.
A comparison between the posterior PDFs for age and extinction obtained for a constant versus variable IMF
(the purple and blue lines respectively) reveals excellent agreement between the two. Besides validating our procedure,
the fact that the age and extinction PDFs computed in these two cases are virtually indistinguishable implies that
\imfs, $A_{\rm V}$, and $T$ have modest covariance, as suggested by the shape of the joint PDFs
in Figure \ref{fig:cl56nhak15} and Figure \ref{fig:cl591nhak15}. 

A similar conclusion holds for the posterior PDF of the mass, although differences between the constant
and variable IMF cases become more noticeable. Indeed, there is some divergence both at very low and very high masses,
particularly with the variable IMF case being skewed on average towards low masses. The origin of this shift,
albeit modest, can be attributed to the degeneracy between mass and IMF highlighted above.
Indeed, compared to the case of a fixed IMF with $\alpha_3 = -2.3$,
the analysis which includes a variable IMF seems to prefer, on average, lower cluster masses, which are populated
by stars that are on average more massive. This effect is indeed visible in the cumulative
posterior PDF for \imfs, which is skewed towards \imfs\ values that
are shallower than the canonical Salpeter value.

\begin{figure*}
  \centering
  \begin{tabular}{cc}

    \includegraphics[scale=0.55]{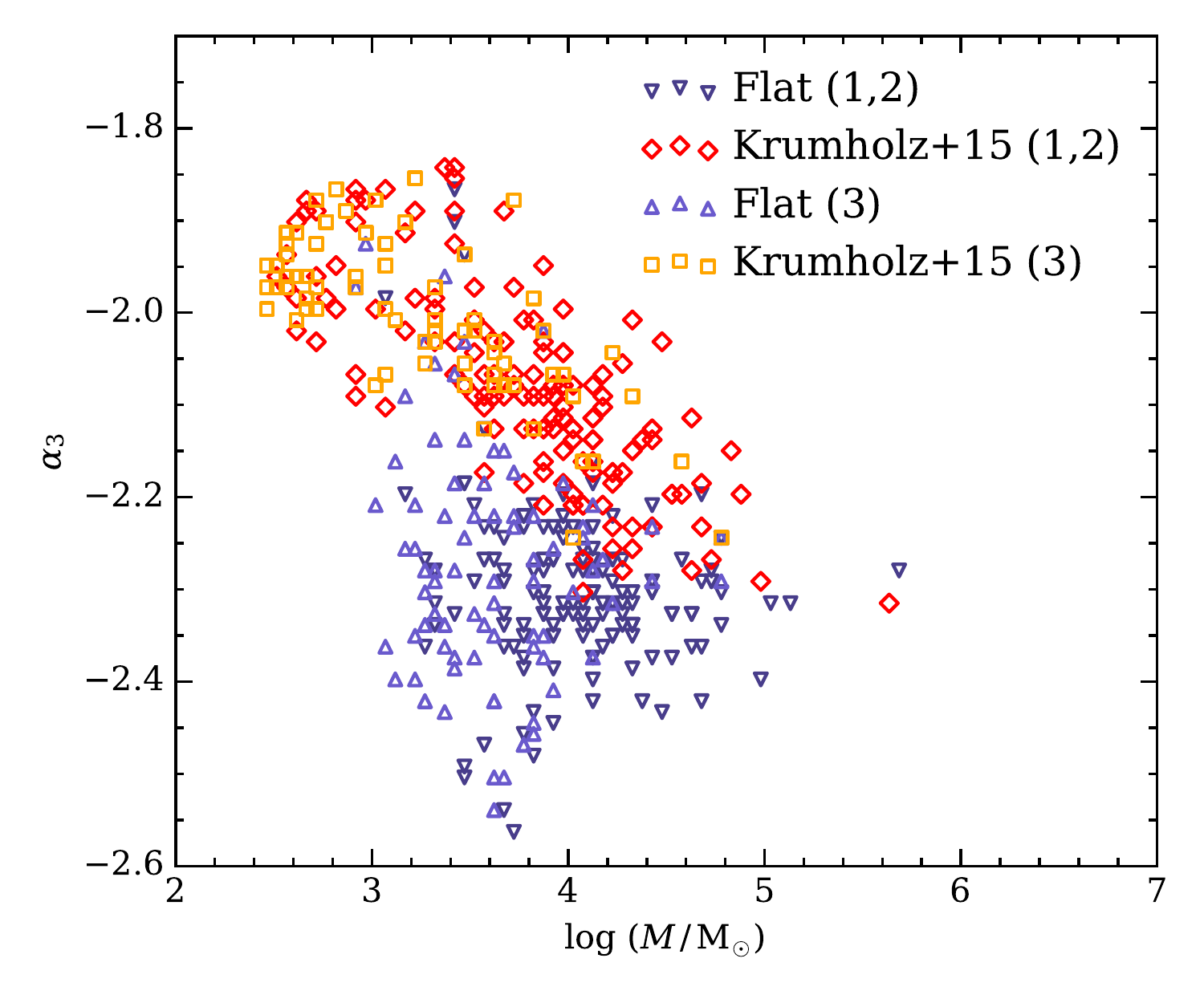} &
    \includegraphics[scale=0.55]{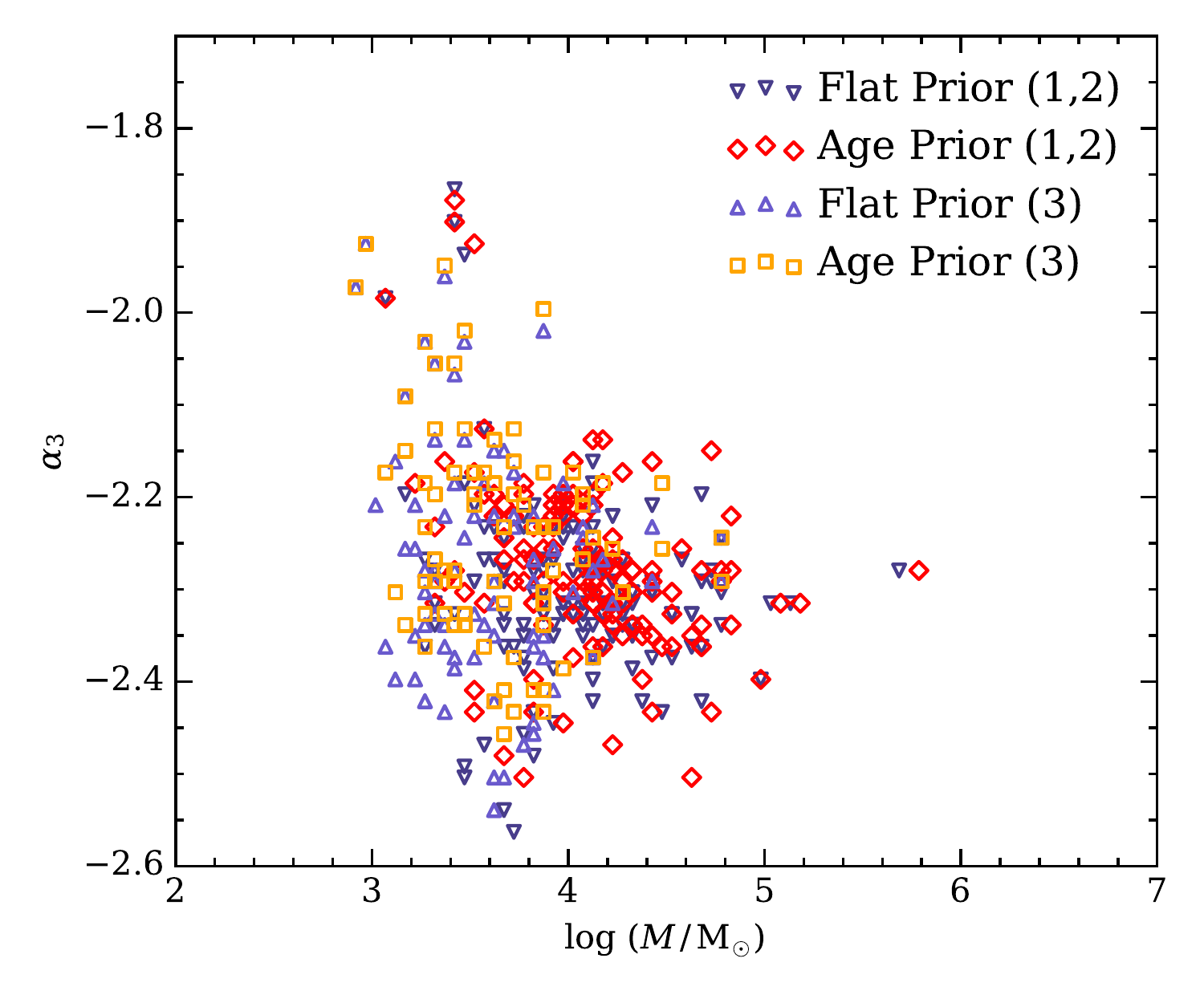}\\
    (a) Physically motivated priors  & (b) Age prior only \\[6pt]    
  \end{tabular}
  \caption{ Panel (a) shows the correlation between the medians of the posterior PDFs for $\log M$ and \imfs, with class 3 objects (light colours) and class 1-2 objects (dark colours)  separated for both flat (triangles) and the \protect\cite{Krumholz2015-SLUGandLEGUS} priors (quadrilaterals). Note the more apparent correlation between mass and \imfs\ for the physically motivated priors, with a lower limit of \imfs\ around the canonical Kroupa value (-2.3). Panel (b) shows the same but with the prior on the mass removed, leaving only a prior on age. This results in a distribution very similar to a purely flat prior.}
  \label{a3vsm}
\end{figure*}

\begin{figure}
\includegraphics[scale=0.4]{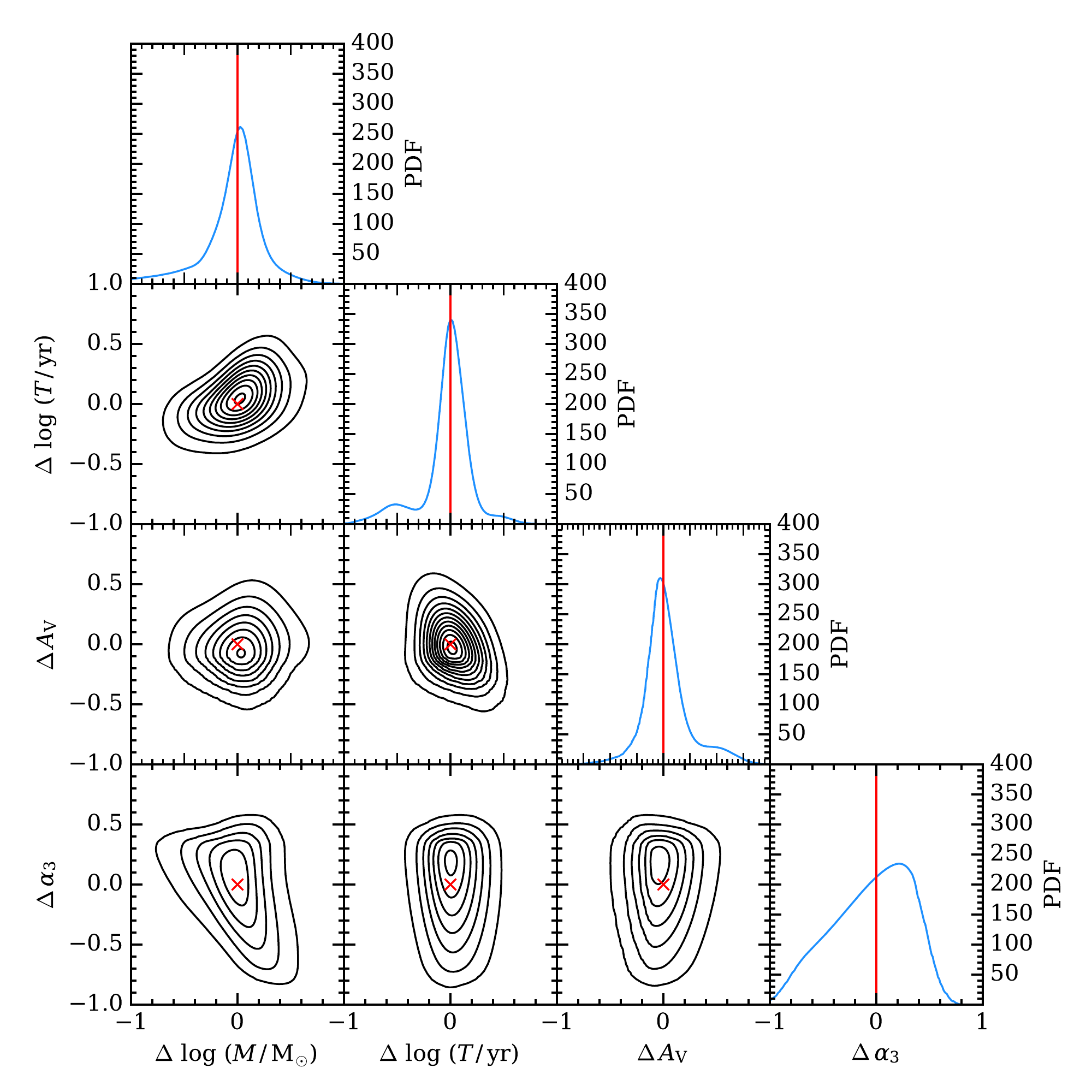} 
\centering
\caption{Stacks of the one and two-dimensional PDFs (median-centred) for the clusters in NGC~628E. The 2D PDF contours are spaced in steps of 50, beginning at 50. The red crosses and lines mark the zero points.}
\label{2dstacks_legus}
\end{figure}

Moreover, Figure \ref{cumulatives} shows that, in line with what is found when examining the PDFs for individual clusters, the
cumulative posterior PDF for \imfs\ is quite broad, implying that only weak constraints on the slope of the IMF can be
obtained when comparing broad-band photometry with simulated clusters. As noted above, however, the posterior
PDF appears skewed towards a shallow IMF, with $\alpha_3 \gtrsim -2.5$. This effect is modest, and
not particularly statistically significant. We can nevertheless inspect the median \imfs\ 
recovered for each cluster, and study whether systematic trends can be found as a function of mass and
LEGUS class. This analysis is shown in Figure~\ref{a3vsm}.
When considering the physically motivated priors (squares and diamonds), a noticeable correlation with mass
appears. In this figure,  we have also separated clusters by their LEGUS class. We see that the majority of the class 3
objects (squares), which include clusters with multiple peaks and what are likely to be associations, 
are recovered at low masses (see also Adamo et al. (2017, submitted)). These objects are characterised by
shallow slopes of the upper-end of the IMF. Conversely, class 1 and class 2 clusters (diamonds) are on average more
massive, although a fit through the data  still indicates a correlation between the median \imfs\ and the cluster mass.

The correlation between the value of \imfs\ and the cluster mass can be explained as follows. Due to the degeneracy between mass and IMF,
the observed light can be modelled either by a relatively massive cluster with
canonical $\alpha_3 = -2.3$, or by a lower mass cluster with a more shallow IMF.
The choice of a physical prior of the form $\propto M^{-1}$ skews the mass PDF towards low values (see also Section \ref{sec:priors}), making the latter case (lower cluster mass and hence a more shallow IMF) preferred.
Therefore, the combination of imposing a physically motivated prior with a varying upper-end slope of the IMF
moves the clusters in the $\alpha_3/M$ plane along a diagonal in the direction of lower mass and
shallower slopes compared to the canonical Kroupa IMF.
From a physical point of view, correlations between the cluster mass and the IMF slope are expected. For instance, a truncation of the IMF in excess of the effect induced by random sampling is predicted by the IGIMF theory, in which low mass clusters have a bottom-heavy IMF
\citep{Kroupa2001-IMF,Kroupa2003-IGIMF,Weidner2011-IGIMF,Kroupa2013-StellarSubStellar}.
However, the effect we see in panel (a) of Figure \ref{a3vsm} is in the opposite direction to the effect predicted by the IGIMF.
The fact that the posterior PDFs for individual clusters are very broad and that,
as discussed below, the median values are quite sensitive to the choice of prior cautions against
far-reaching conclusions of the nature of this correlation. Indeed, at this time we cannot exclude the possibility that the observed trend arises from a mass-dependent
correlation at second order that is not explicitly captured in our analysis. 

\begin{figure}
\centerline{\includegraphics[scale=0.425]{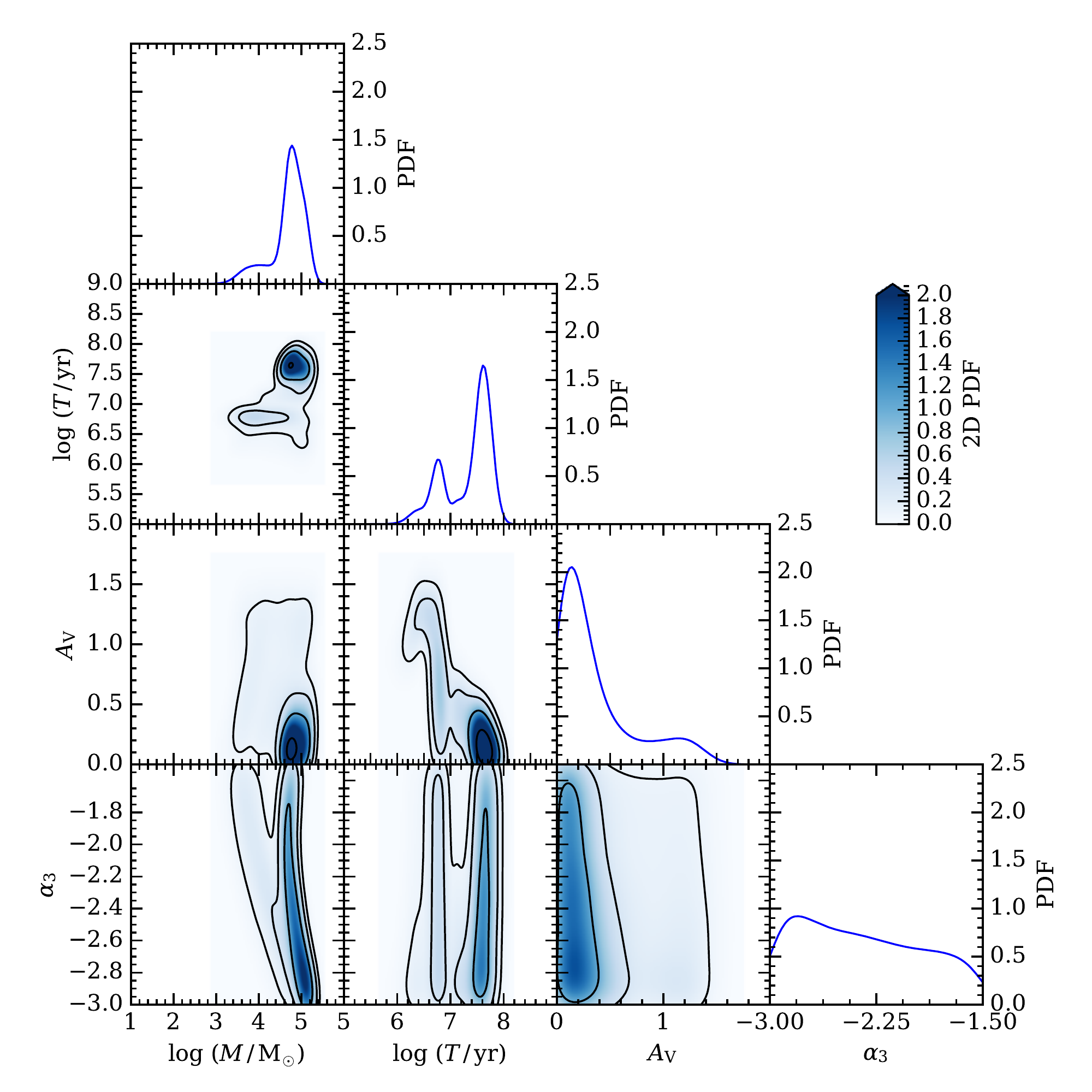}}
\centering
\caption{As Figure \ref{fig:cl56nhak15}, but with a flat prior on all physical parameters.
}
\label{fig:cl56nha}
\end{figure}

\begin{figure}
\centerline{\includegraphics[scale=0.45]{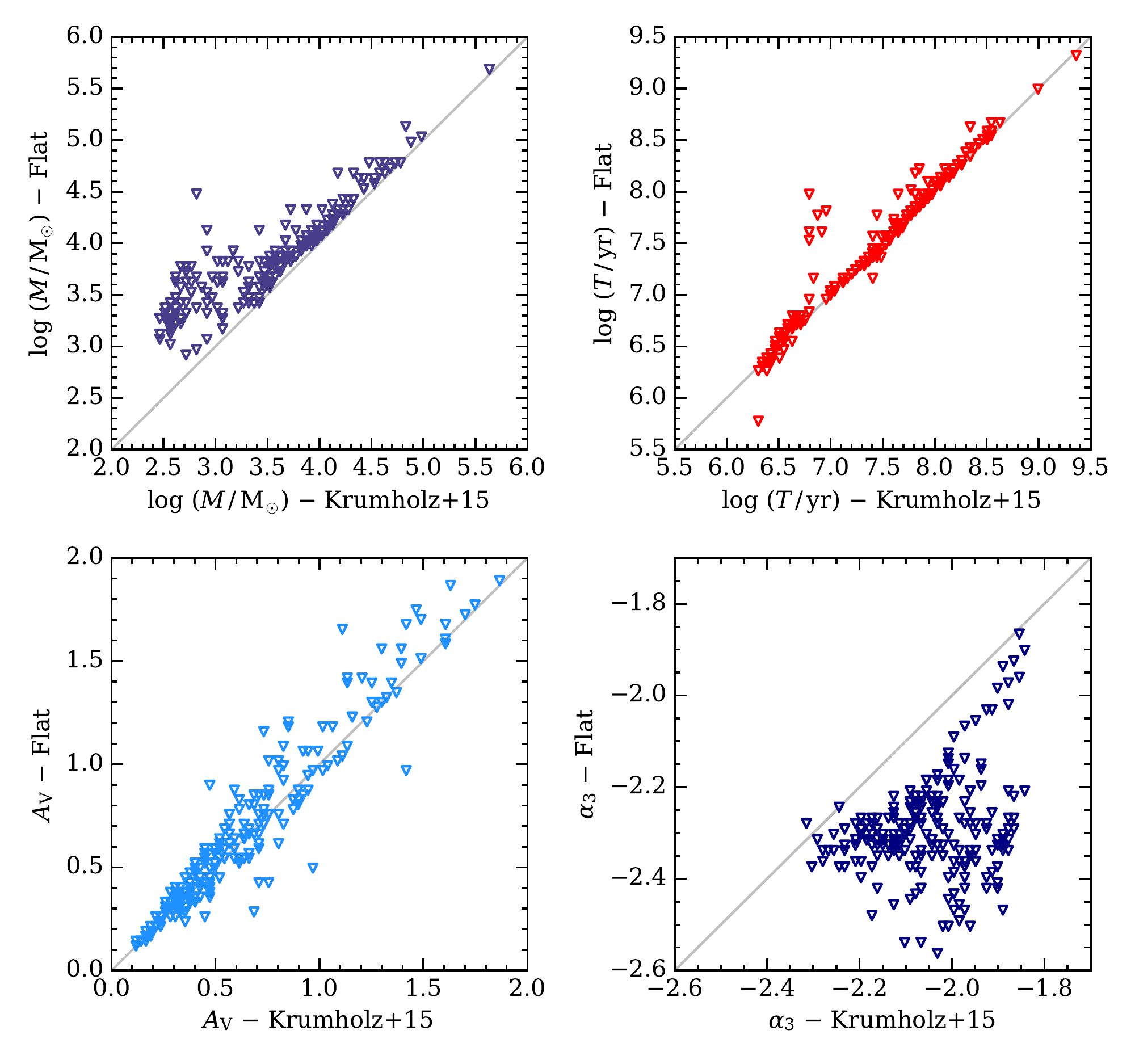}}
\centering
\caption{Comparison between the median values for each parameter returned by \clus\ for flat priors and the physically motivated
  priors by \protect\cite{Krumholz2015-SLUGandLEGUS}. The grey line marks the one-to-one correspondence. While extinction and age are
  insensitive to the choice of priors, mass and $\alpha_3$ depend on the assumed prior, owing to a degeneracy between these two parameters.}
\label{pvsp}
\end{figure}

Finally we produce a median-centred stack of the 1D and 2D posterior PDFs for the LEGUS clusters, which is shown in Figure~\ref{2dstacks_legus}. Degeneracies between the cluster mass and age, and between the cluster mass and IMF slope (as seen in panel (a) of Figure~\ref{a3vsm}) are apparent from the shape of the contours in their respective joint posterior PDFs. The broad shape of the posterior PDFs in \imfs\ are also visible. The mass-age degeneracy is, at least in part, caused by the need for a higher mass to produce the same luminosity in older clusters.

\subsection{Effects of the choice of prior}\label{sec:priors}

The analysis of clusters from NGC~628E using the \clus\ tools presented in the previous section shows that age and extinction
are largely insensitive to the parameters describing the IMF. Moreover, we have shown that the posterior PDF of
\imfs\ is weakly constrained by broad-band photometry. When choosing a physically motivated prior on mass,
the recovered PDFs are skewed towards lower masses and shallower IMF slopes compared to the case of a constant
Kroupa IMF. As shown above, this effect arises from a degeneracy between mass and \imfs, which results in
diagonal shifts in the $M/\alpha_3$ plane.
It is therefore expected that the choice of prior on the mass  will have a consequential effect on the parameters
describing the IMF (and vice versa), as we show in this section.

We start by examining again the corner plot for the example cluster ID~56, this time computed assuming a flat prior on all the parameters,
which is shown in Figure \ref{fig:cl56nha}. A comparison with Figure \ref{fig:cl56nhak15}, in which a physically motivated
prior was chosen, highlights how a flat prior shifts the probability from
a locus of low mass and young age ($M\sim 10^{3.5}~\rm M_\odot$, $T\sim 10^{6.7}~\rm yr$)
to a locus of higher mass and age ($M\sim 10^{5}~\rm M_\odot$, $T\sim 10^{7.6}~\rm yr$).
As a consequence, in our model based on stochastic sampling, the observed photometry can be realised at higher
cluster masses with steeper IMFs ($\alpha_3\lesssim -2.3$) compared to the case of low cluster mass.
In other words, to produce a comparable UV luminosity to that of a higher mass cluster, a low mass cluster needs a larger number
of massive stars, something that can be achieved with a more shallow IMF slope.

By repeating the analysis for the full sample but using a flat prior, we see that the behaviour described above is
indeed general. Compared to a physically motivated prior, a flat prior induces a shift in the cumulative PDFs
(red lines in Figure \ref{cumulatives}) towards higher masses and steeper IMF slopes. Conversely, there is no noticeable
difference in the extinction and only a modest change occurs in the age cumulative distribution, with some probability flowing
from $T \gtrsim 10^{6.5}~\rm yr$ to $T \lesssim 10^{6.5}~\rm yr$. As a consequence of this shift in mass/\imfs, the correlation
between the cluster mass and the IMF slope disappears when considering a flat prior (triangles in panel (a) of Figure \ref{a3vsm}). 

Panel (b) of Figure~\ref{a3vsm} shows the same plot but with the analysis performed using only the age component of the \cite{Krumholz2015-SLUGandLEGUS} prior. Here we see a distribution of points that is almost identical to that of a flat prior, suggesting that the mass prior is dominant in its effect on the IMF that we recover. On these
plots, we note that the fact the medians are centred around $\alpha_3 \approx -2.3$ is not indicative that the data prefer a Kroupa IMF.
Rather, this is a mere reflection of the fact that the posterior PDFs on the IMF slope are very broad and, by construction,
are centred around the canonical value of $\alpha_3 \approx -2.3$. Based on this result, it may be tempting to conclude that
the choice of a flat prior is preferable over the physically motivated prior of \cite{Krumholz2015-SLUGandLEGUS}. However,
while we do not wish to over-interpret the trend visible Figure \ref{a3vsm}, 
the choice of our physically motivated prior is preferable given the power-law nature of the cluster
mass function. This implies that either our \slug\ simulations are not fully describing the properties of clusters in
NGC~628E, or that indeed there is (weak) preference for more shallow IMF slopes in low mass clusters. 

Finally, rather than considering the cumulative PDFs, we compare the medians of the posterior
PDFs for the physical parameters of individual clusters, as shown in Figure \ref{pvsp}.
Consistent with the previous discussion, the medians of the age and extinction PDFs
are insensitive to the choice of prior, while the medians for the mass are systematically offset in the
direction of lower masses for a physically motivated prior. This shift is modest ($\sim 0.2~$dex)
for clusters above $\sim 10^{3.5}~\rm M_\odot$, but it becomes more significant (up to $\sim 0.5~$dex)
at the low mass end. This figure also shows that shifts of a factor of 2 in mass that are
modest on a logarithmic scale result in more appreciable variations of a power-law
exponent on a linear scale, with $\Delta \alpha_3 \sim 0.2-0.3$. 

One possible way to avoid the IMF slope having such a dependence on the mass prior would be to analyse all the clusters as a single unit with a common mass function between them. At present, a single cluster is considered more likely to be less massive, as suggested by our choice of prior. However, for an ensemble of clusters which follow the CMF that forms the prior on the mass, we would retrieve a distribution of masses that follows this CMF. In this situation there would not be a preference for every cluster to be less massive, rather for the ensemble of clusters to have a preference for less massive clusters.

\begin{figure*}

 \centering
  \begin{tabular}{cc}

    \includegraphics[scale=0.425]{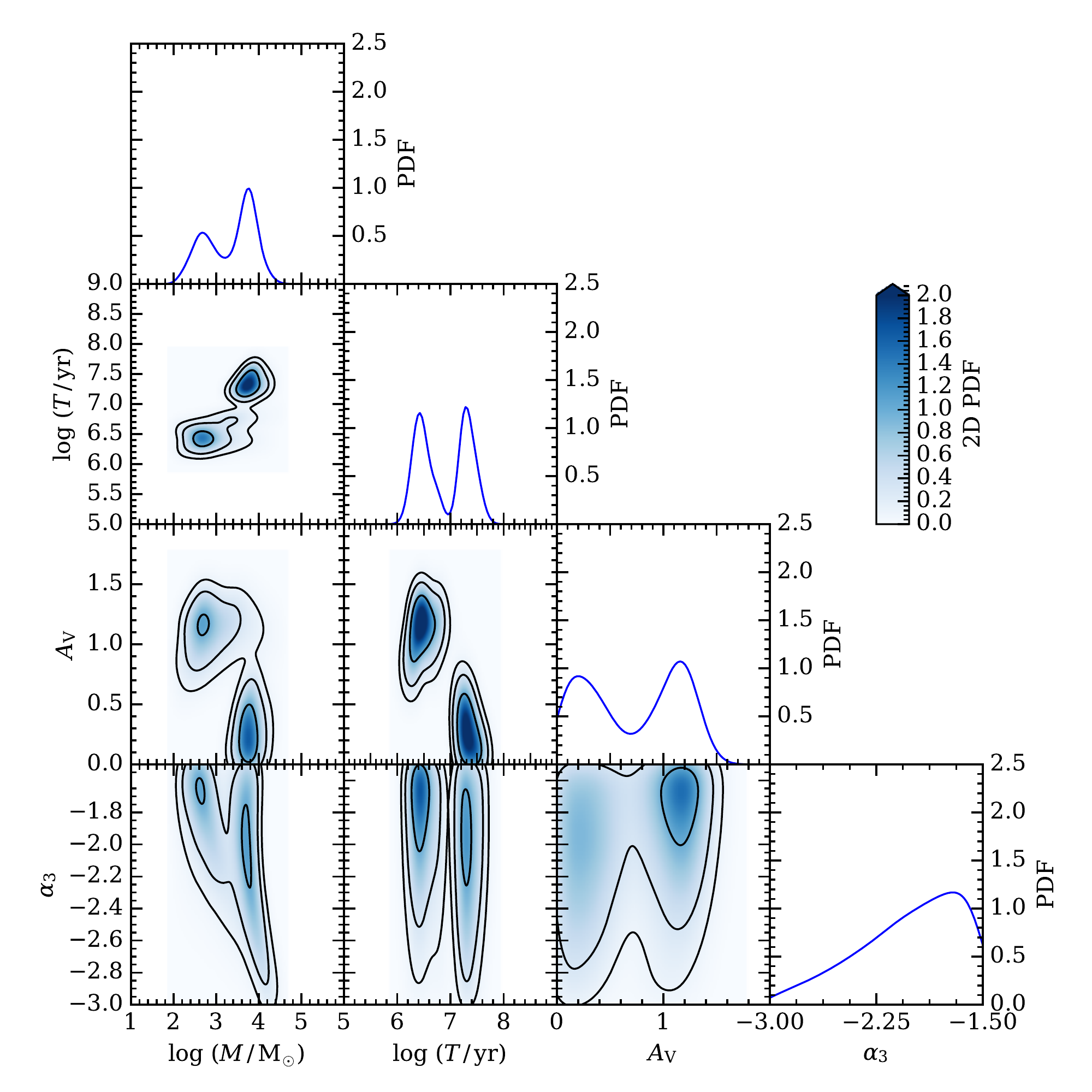} &
    \includegraphics[scale=0.425]{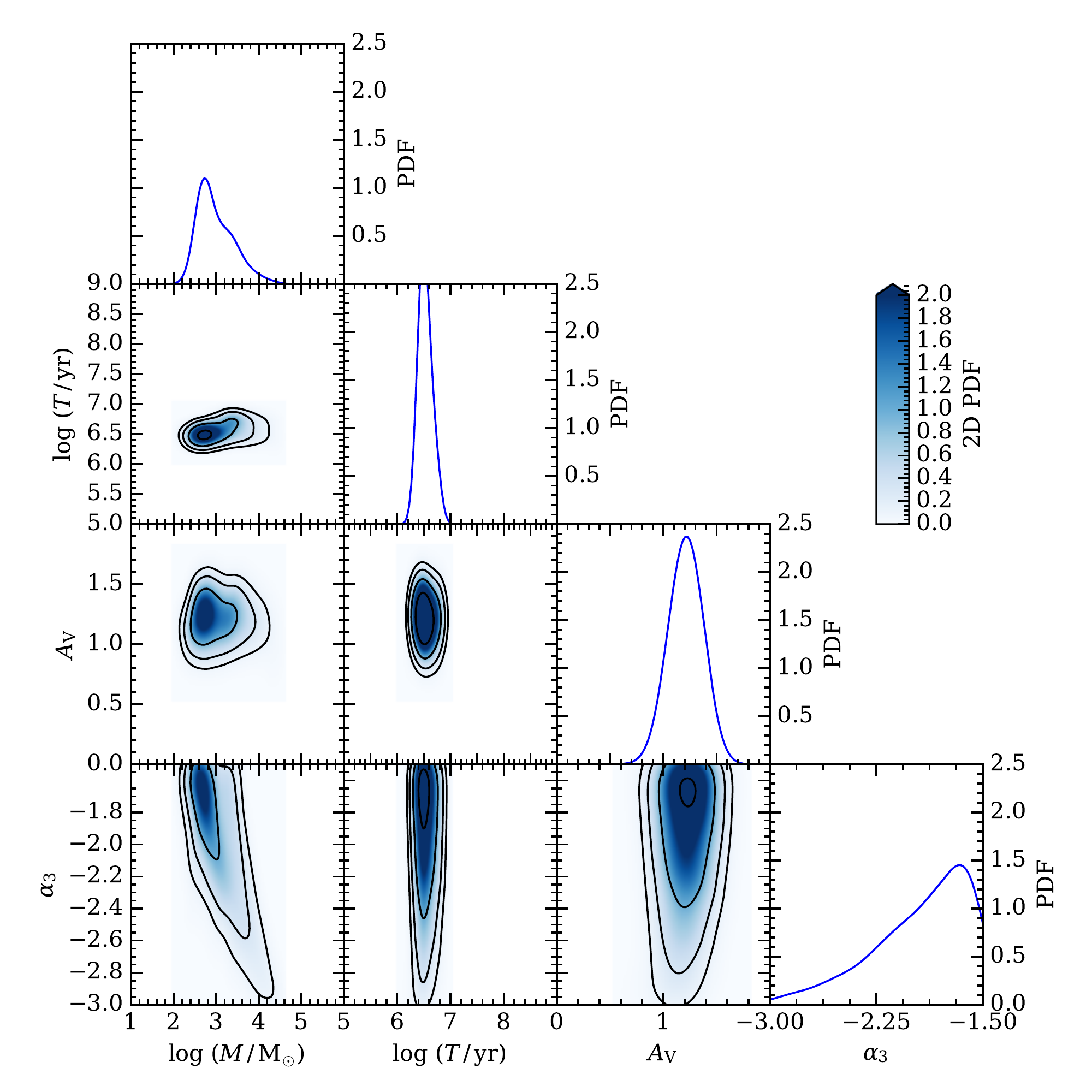}\\
    (a) Broad-band filters only  & (b) Broad-band + H$\alpha$ \\[6pt]    
  \end{tabular}
  \caption{As Figure~\ref{fig:cl56nhak15}, but for cluster ID 292, which has observed H$\alpha$ emission.
      Panel (a) is for the standard LEGUS broad-band filters only, whereas panel (b) includes the F657N filter to represent
      \ha. We see significant improvement in the recovery of the mass, age, and extinction.}

\label{fig:cl292hak15}
\end{figure*}

\begin{figure*}
  \centering
  \begin{tabular}{cc}

    \includegraphics[scale=0.425]{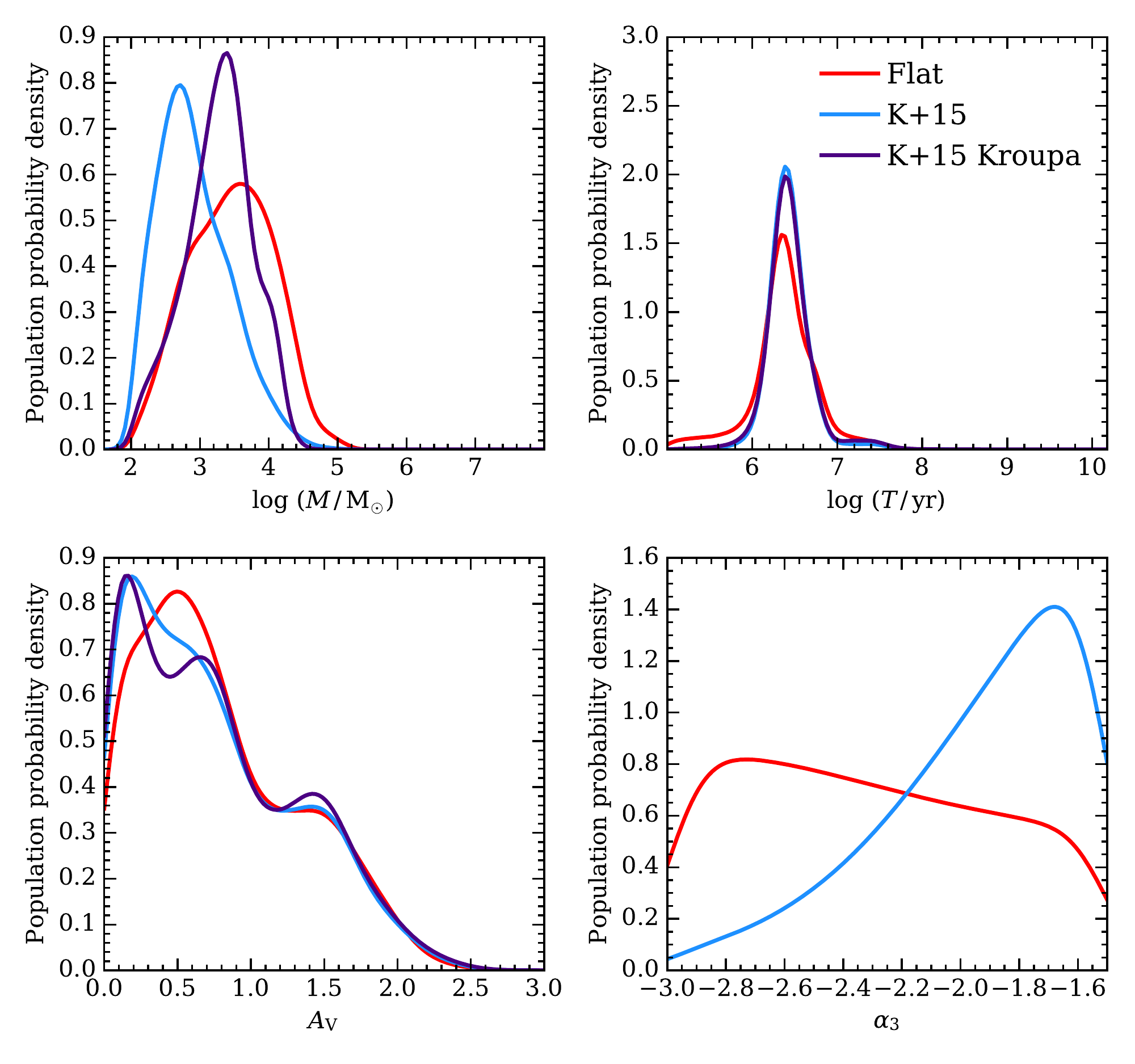} &
    \includegraphics[scale=0.425]{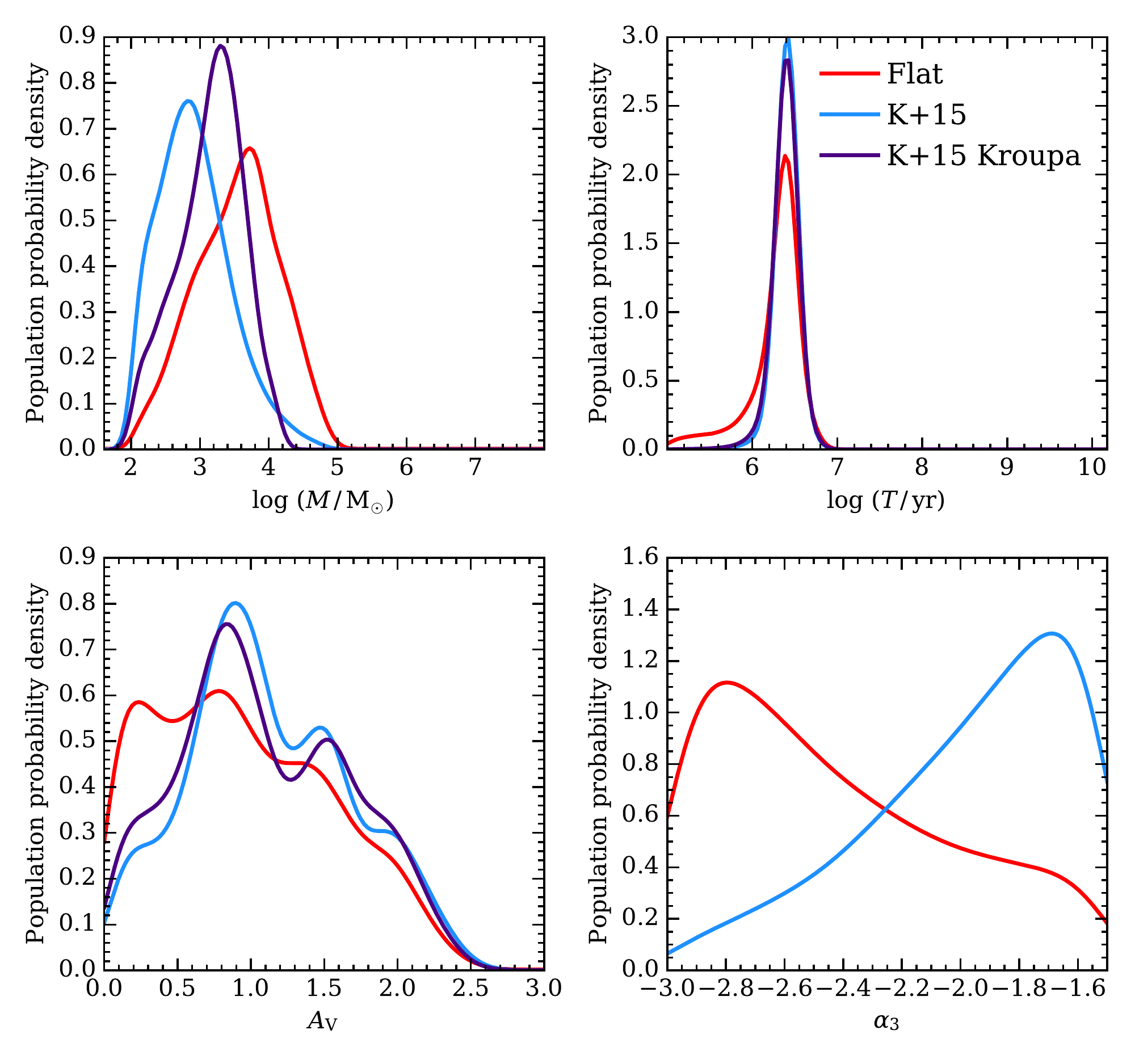}\\
    (a) Broad-band filters only  & (b) Broad-band + H$\alpha$ \\[6pt]    
  \end{tabular}
  \caption{The cumulative posterior PDFs for a subset of young clusters (with median age $T \leq 10^{6.5}~\rm yr$)
    computed for both choices of priors, and for a constant Kroupa IMF. The left panels show the PDFs
    based on broad-band photometry only, while the right panels show the PDFs computed including the H$\alpha$ flux. The inclusion of \ha\ in the analysis results in much sharper PDFs in age, which also lifts some of the degeneracy in extinction. There is also a small shift towards steeper IMF slopes for both priors.}
  \label{cumulatives_ha_young}
\end{figure*}

\subsection{Adding H$\alpha$ photometry to the Bayesian analysis}\label{sec:halegus}

\subsubsection{Preparation of the H$\alpha$ photometry}\label{sec:haprep}

As shown in Section~\ref{sec:legusanalysis}, our analysis of star clusters from NGC~628E
based on the five broad-band LEGUS filters provides only modest constraints on the IMF slope. We also find that
several clusters have broad or multi-peaked posterior PDFs in
some of the physical parameters, especially age. As massive OB stars dominate the \ha\ emission in clusters at younger
ages, the inclusion of \ha\ photometry is expected to improve some of the constraints on the physical parameters of interest.
To this end, we include in our analysis data obtained as part of H$\alpha$-LEGUS (Chandar et al., in preparation) that are
collected through the WFC UVIS F657N filter which, at the redshift of the LEGUS galaxies, encompasses the \ha\ line.

To derive aperture photometry for \ha\ data, we follow the method developed by Lee et al. (in preparation), which we briefly summarise here.
During the processing of the F657N data, the standard approach for applying aperture corrections to broadband photometry
  cannot be trivially applied to the nebular emission, given that the stellar continuum and the nebular emission
  differ in their spatial extent.  To account for this difference and to produce photometry that can be compared to \slug\ simulations,
  we  follow a two step procedure.
  
Firstly, we synthesize an aperture-corrected continuum flux in the F657N filter, $f_{\rm 657,c}$.
This is achieved by linearly interpolating the aperture-corrected continuum flux measured in the
F547M and F814W filters, with a weight tuned to recover the observed aperture-corrected flux in the F657N filter
for clusters without H$\alpha$ emission.
This procedure yields a synthesised F657N aperture-corrected continuum flux.
Next, we repeat the above procedure, but for fluxes that are not aperture corrected, thus obtaining an interpolated 
continuum flux in the F657N filter that is not aperture-corrected. This is needed to compute the line emission, as explained next.
To derive the net \ha\ line flux $f_{\rm 657,H\alpha}$, we then subtract the synthesised continuum flux (without aperture correction)
from the measured flux in the F657N filter.

Combining the two steps, we produce the aperture-corrected total flux (continuum plus line) in the
F657N filter as $f_{\rm 657,corr} = f_{\rm 657,H\alpha} + f_{\rm 657,c}$.
Thus, at the end of this procedure, we have created a synthetic aperture
correction from the weighted average of the aperture corrections on the bracketing filters and applied this to the measured F657N flux.
Finally, as for the LEGUS broad-band photometry, we apply a correction for Galactic extinction. This results in a total flux value for the F657N filter which can be used in our analysis and is comparable to the F657N photometry generated by \slug.
This total flux (hereafter referred to H$\alpha$ flux for simplicity) is the quantity that we compare directly with
the output of \slug. 

The SPS module at the core of the \slug\ package predicts the number of ionisation
photons produced by stars in each cluster, which we convert into nebular emission
following the methods described in \citet{Krumholz2015-SLUGiii}. At this stage, we assume that only a
fraction ($\phi = 0.5$) of the ionising photons are converted into nebular emission. After convolving
the simulated cluster SED with the F657N filter transmission curve, \slug\ generates a total flux like
the $f_{\rm 657,corr}$ we synthesise for the observations as described above. 
Throughout this work, we do not include H$\alpha$ emission from stars themselves (e.g. from Be stars),
as they are found to contribute only a few percent compared to the nebular emission.

We note that, due to the extended nature of the {\ha} emission, there is a possibility that we are missing some
  fraction of the flux in the F657N filter from extended (but typically low surface brightness) emission.
  As we will show below, however, we are able to reproduce all the findings of our analysis using a set of mock
  clusters that, by definition, are not affected by loss of flux on large scales. This implies that any potential
  loss of flux has a minor effect on our results.
  Moreover, the choice of a nebular fraction of $\phi=0.5$ already mitigates this effect.
  We have also conducted tests to verify that our results are not sensitive to the exact choice of this coefficient.
  Indeed, we find that repeating our analysis using a library with $\phi=0.7$  does not have a noticeable
  effect on the medians of the PDFs we return for the physical parameters. 

\subsubsection{Results including H$\alpha$ photometry}

To illustrate the effect of including H$\alpha$ in our analysis, we turn to an example cluster
  with observed H$\alpha$ emission (ID 292), which is
  shown in Figure \ref{fig:cl292hak15}. Looking at the differences between panels (a) and (b),
we see a significant improvement in the age determination, with the locus at $T \sim 10^{6.5}~\rm yr$ being
clearly preferred over the older ages. In this particular case, the inclusion of H$\alpha$ further improves the
constraints that can be derived on the mass and extinction, removing the bimodality seen in both.
In turn, as the lower mass at $M\sim 10^3~\rm M_\odot$ is now preferred, the posterior PDF for \imfs\ sharpens
compared to the case in which no H$\alpha$ was included. 

As we will show in the following section, the improvement in age determination obtained when including \ha\ in the analysis is common to much of the cluster sample, and is most noticable in the case of younger clusters. However, the variation in the median of the age PDFs is modest, and we see no appreciable difference in the shape of the cumulative PDFs when considering our entire cluster catalogue. Thus, given that the shapes of the cumulative PDFs for both mass and age are primarily dependent on the scatter among clusters rather than the width of the PDFs for the individual clusters, the shapes of the PDFs shown in Figure~\ref{cumulatives} do not show significant change following the inclusion of \ha.
Similarly, we see no appreciable difference in the cumulative PDF for \imfs. This is primarily caused by the broad shape of the \imfs\ PDFs in the individual clusters, even following the inclusion of \ha.

Finally, we examine the effect of including \ha\ in the analysis of young star clusters by limiting our choice of clusters to those whose posterior PDF in age has a median of $T \le 10^{6.5}~\rm yr$. It is for this subset of clusters that the NUV and \ha\ photometry of the LEGUS and \ha-LEGUS surveys is expected to yield the tightest constraints.
The cumulative PDFs for this subset of clusters, computed with both choices of priors, a fixed Kroupa IMF, and with or without \ha, are shown in Figure~\ref{cumulatives_ha_young}. 
This subset contains 21 clusters for the case of flat priors with no \ha, 24 with \ha, 21 for physical priors with no \ha, 18 with \ha, and 20 for the case of the fixed Kroupa IMF with no \ha\ with 21 when \ha\ is included.

Several interesting features can be seen in this figure. Considering first the PDFs derived assuming the physically motivated
priors (blue curves), the inclusion of H$\alpha$ significantly improves the age constraint (compare left and right panels),
thus sharpening the age posterior PDFs. However we see no appreciable shift in the positions of the medians of the PDFs. A more subtle
difference is visible when comparing the mass and \imfs\ PDFs. Conversely, more substantial
differences appear when examining the posterior PDFs for the extinction, with the inclusion of H$\alpha$ favouring
extinctions up to $\sim 0.5\,$dex higher. Comparisons with the PDFs computed with a constant Kroupa IMF (purple line)
reveal that the differences observed when including H$\alpha$ are independent of the choice of the IMF.

Moreover, in line with the discussion presented in the previous section,  Figure \ref{cumulatives_ha_young}
shows how critical the mass-\imfs\ degeneracy is in this type of analysis. At fixed age,
a physically motivated prior skews the mass determination towards smaller values, which is compensated by
a more shallow slope on the upper-end of the IMF. Conversely, a flat prior (particularly when including H$\alpha$)
prefers steeper IMF slopes which result in higher masses. In between, the constant Kroupa IMF
case with $\alpha_3 = -2.3$ places the preferred masses at intermediate values between those inferred
with flat and physically motivated priors. In summary, a key result of this work is that,
due to the mass-IMF degeneracy, the choice of prior on the mass strongly affects the resultant shape of the posterior PDF
of \imfs, preventing robust constraints on the upper end slope of the IMF even when including H$\alpha$. Turning this argument
around, the assumption of a fixed Kroupa IMF, which is equivalent to imposing a strong prior on \imfs, affects the mass
determination by up to $\sim 0.5~\rm dex$ for young and low-mass clusters.

\input{table_mocks.tex}

\begin{figure}
\includegraphics[scale=0.4]{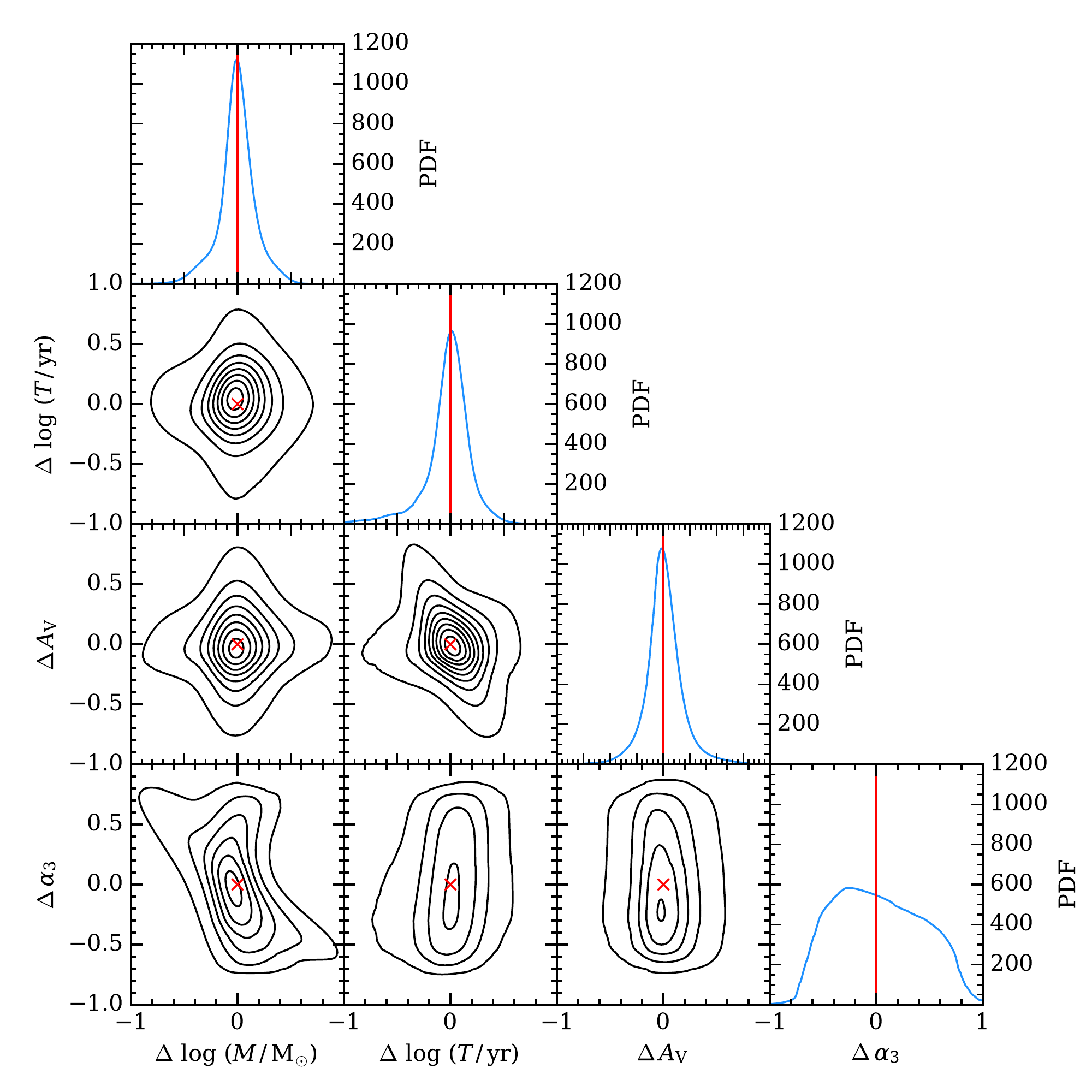} 
\centering
\caption{Stacks of the one and two-dimensional PDFs (median-centred) for 700 mock clusters simulated with a variable IMF and the choice
  of parameters listed in Table \ref{tab:mockparam}. 2D PDF contours are spaced in steps of 100, beginning at 100. The red crosses and lines mark the zero points.}
\label{2dstacks}
\end{figure}

\begin{figure*}
  \centering
  \begin{tabular}{cc}
    \includegraphics[scale=0.425]{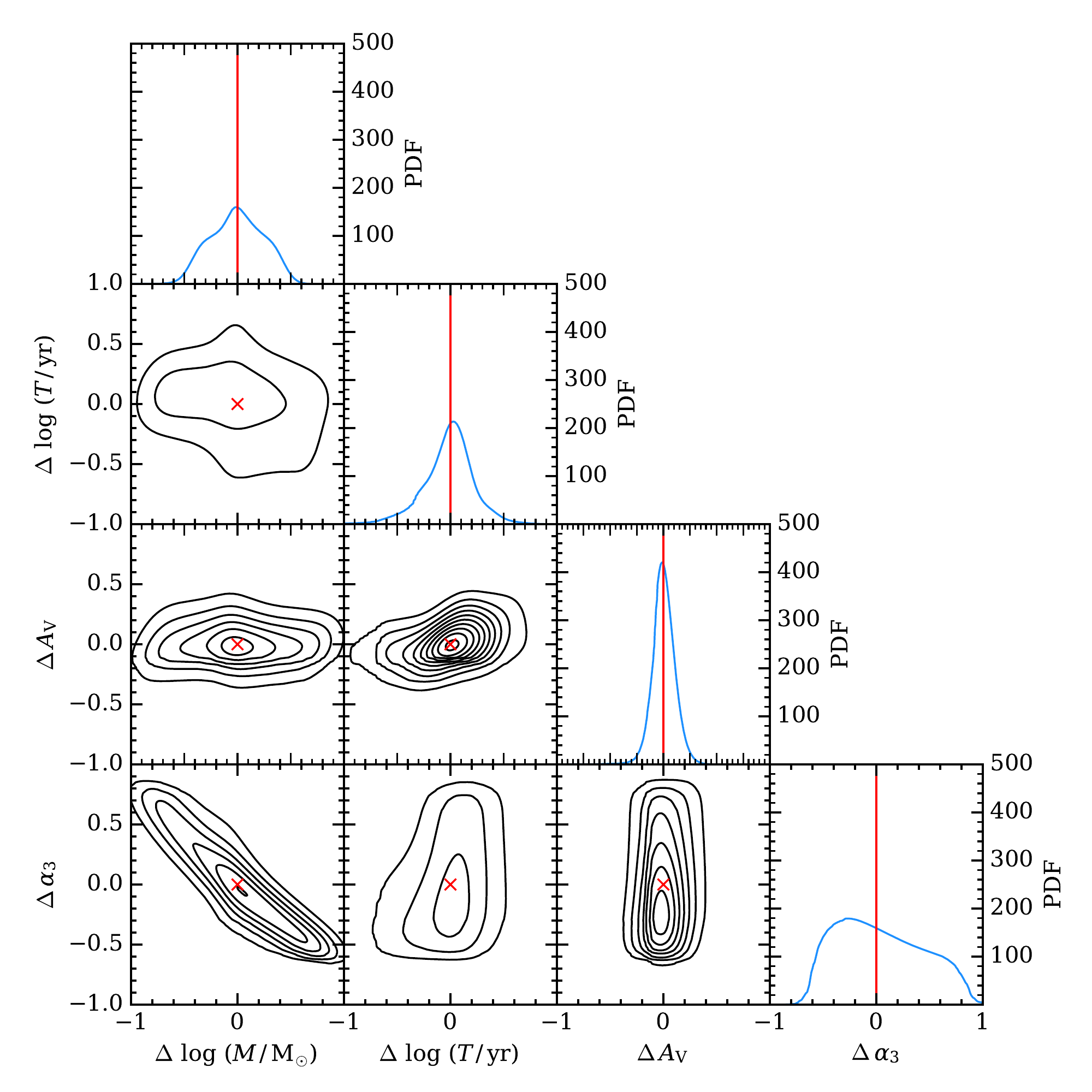} &
    \includegraphics[scale=0.425]{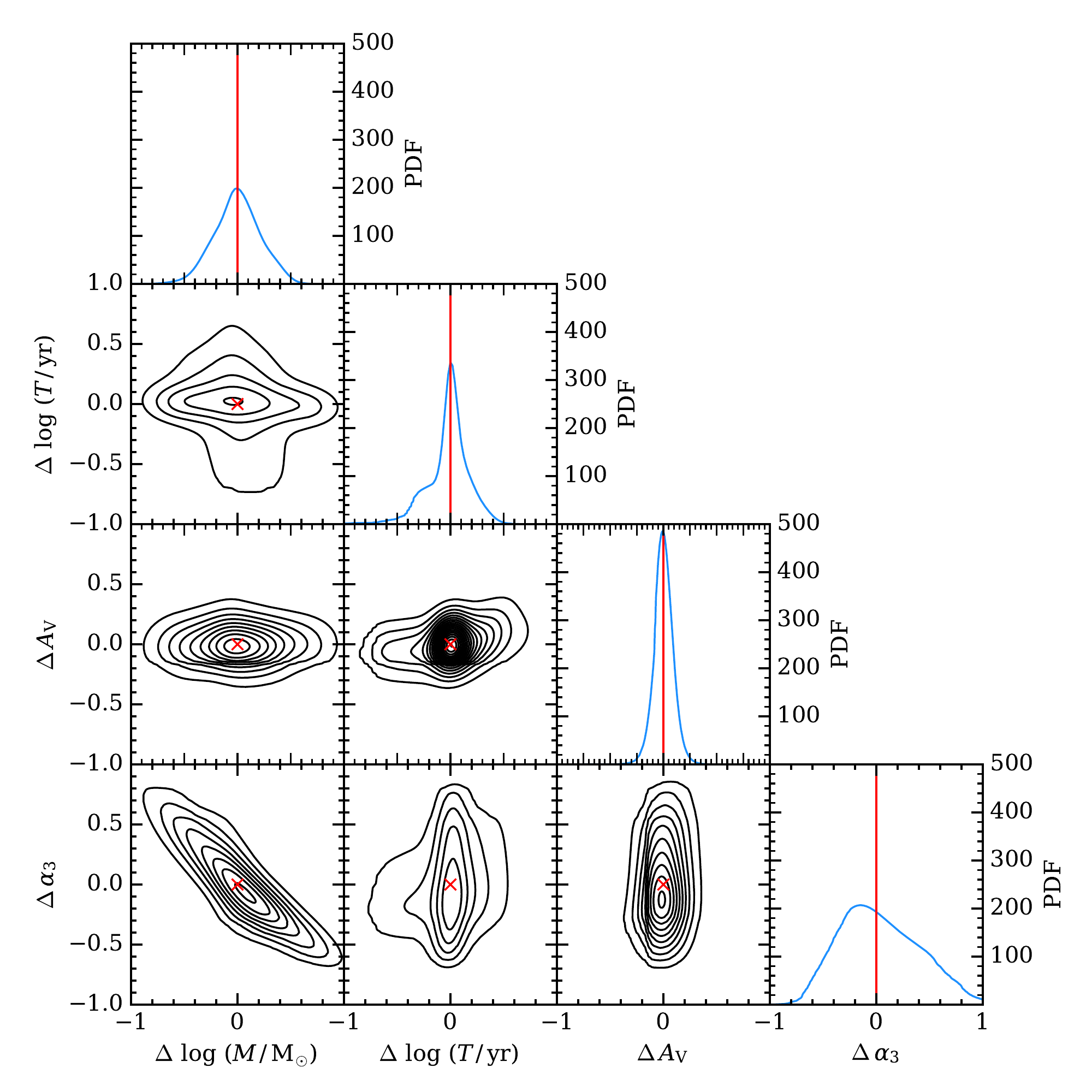} \\
    (a) Broad-band filters only  & (b) Broad-band + H$\alpha$ \\[6pt]

  \end{tabular}
  \caption{As Figure \ref{2dstacks}, but for a subset of young mock clusters, with age $T \le 10^{6.5}~\rm yr$.
    The left and right panel show, respectively, the PDFs recovered from broad-band photometry alone, and
    with the inclusion of H$\alpha$. 2D PDF contours are spaced in steps of 50, beginning at 50. The red crosses and lines mark the zero points.} 
\label{2dstacksYMnHa}
\end{figure*}

\begin{figure*}
\begin{center}
\begin{tabular}{cc}
\includegraphics[scale=0.55]{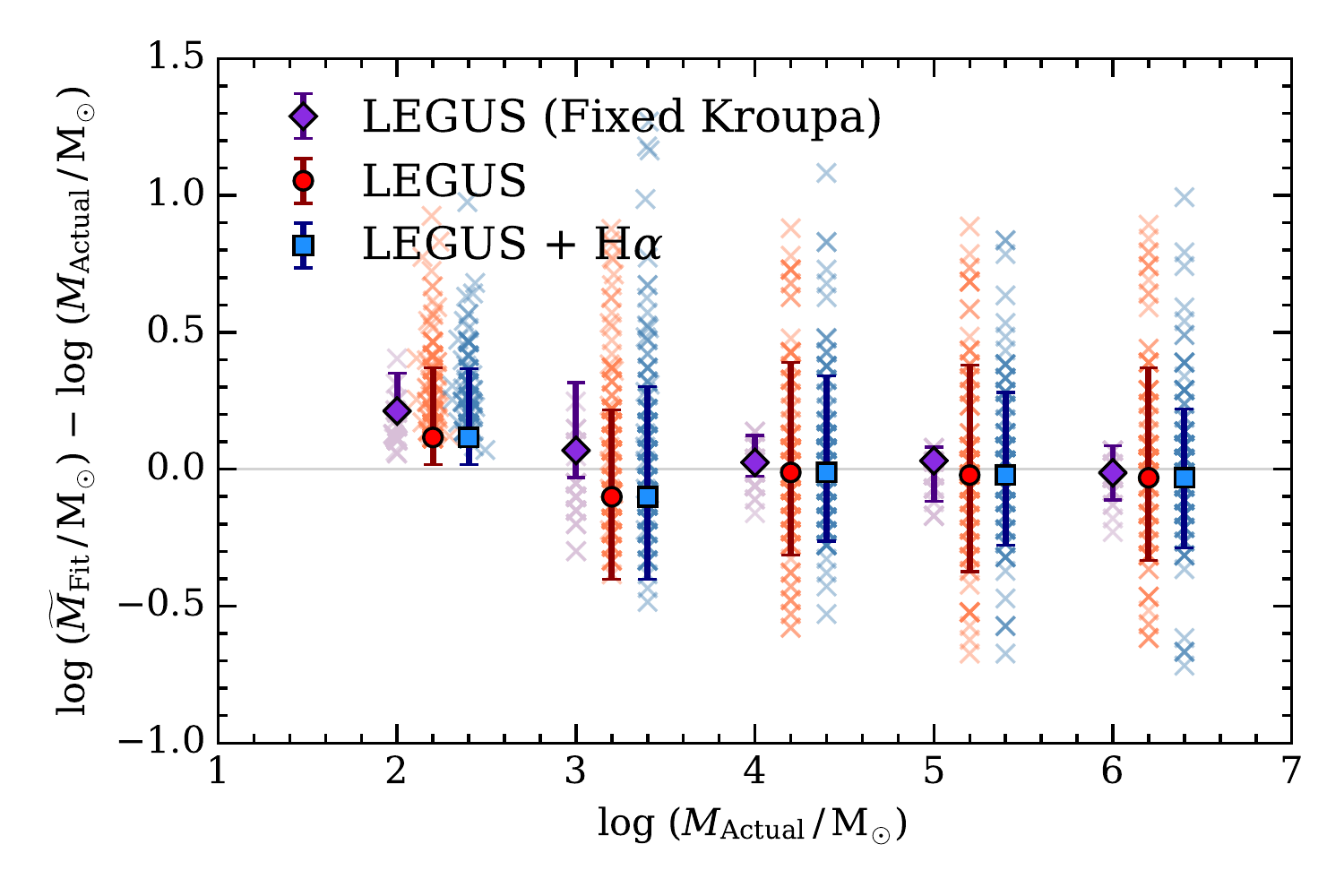} & \includegraphics[scale=0.55]{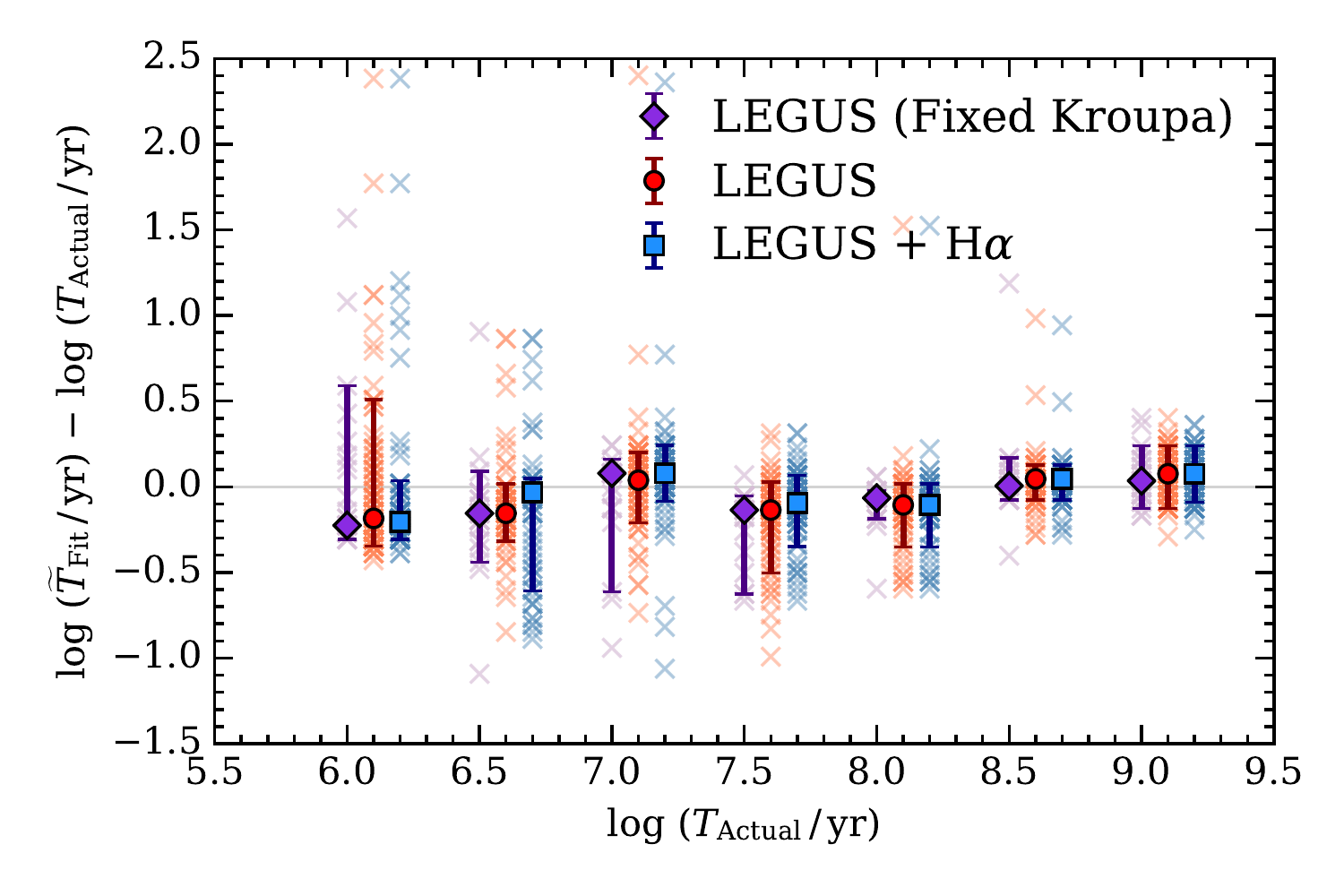} \\
(a) Mass & (b) Age \\[6pt]
\includegraphics[scale=0.55]{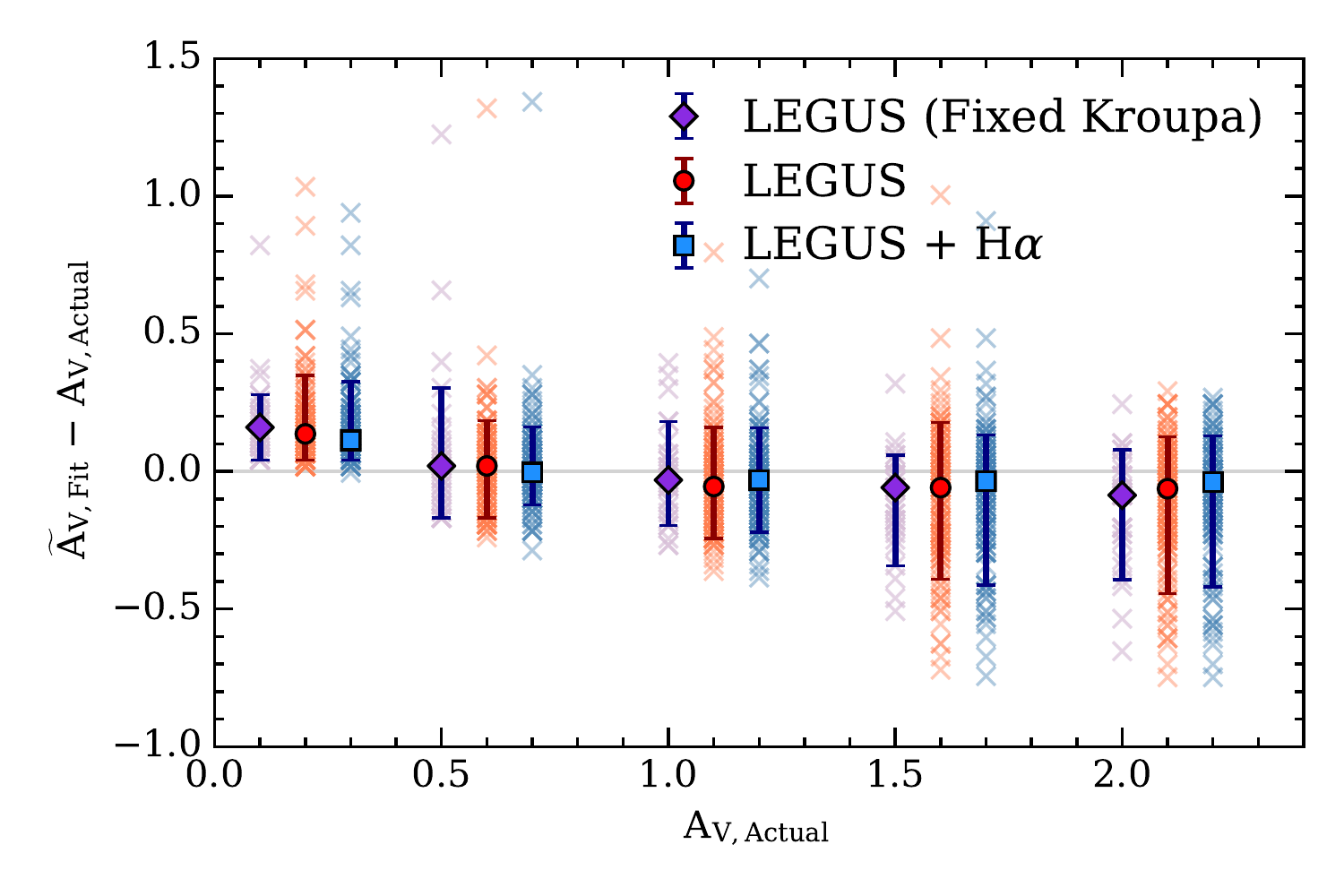} & \includegraphics[scale=0.55]{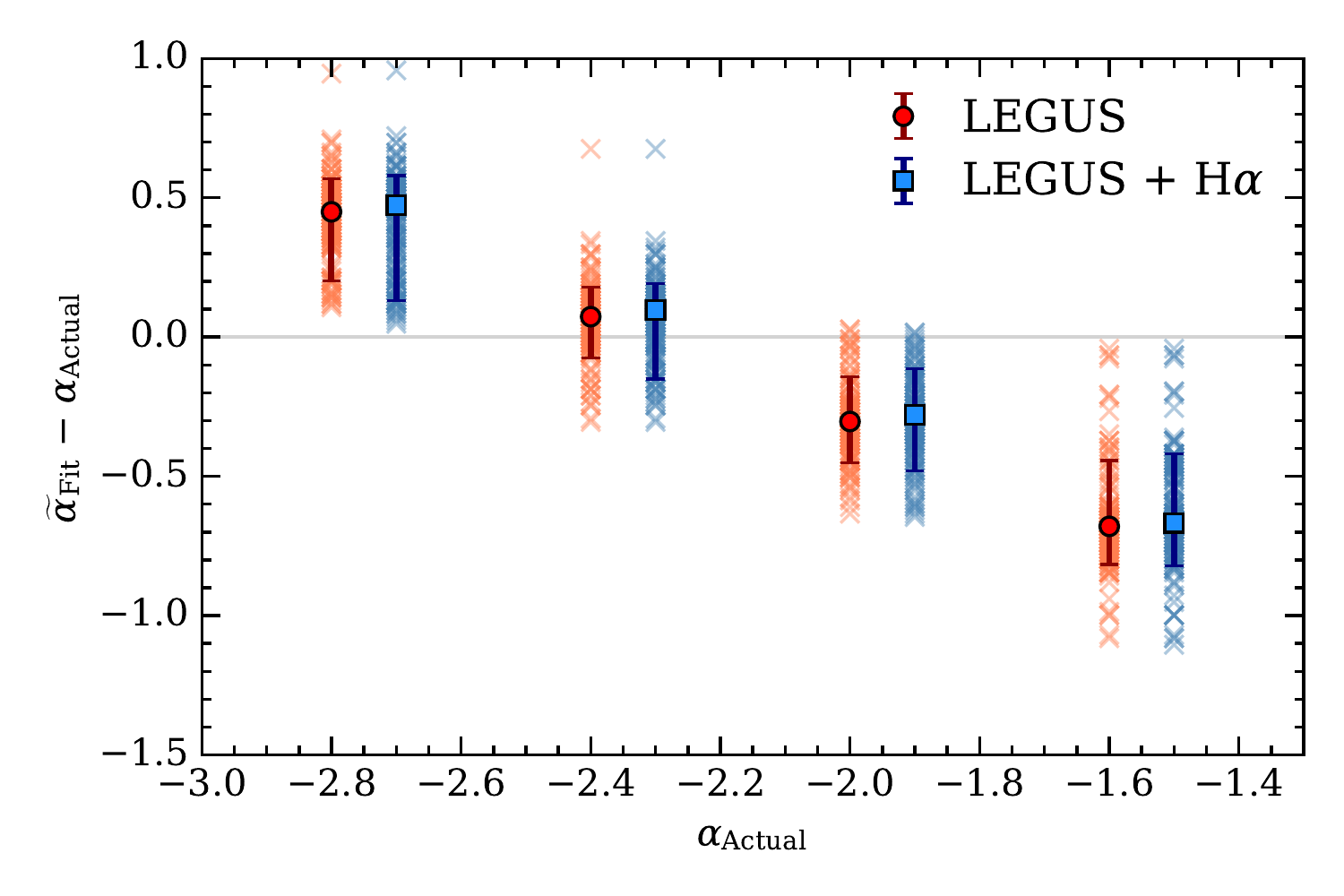} \\
(c) Extinction & (d) IMF \\[6pt]
\end{tabular}
\end{center}
\centering
\caption{Comparisons of the medians of the inferred posterior PDFs for all physical parameters
  to the true values adopted when simulating the mock clusters. Results for individual clusters
  are shown with crosses, while the median residuals and the corresponding $10^{\rm th}$ and
  $90^{\rm th}$ percentiles are shown with solid points. We plot analysis including broad-band photometry alone with (red, circles) and without (purple, diamond)
  a variable IMF. We also show results of the variable IMF case when including H$\alpha$ (blue, squares).
  Points are shifted along the x-axis for visualisation purposes, with the leftmost point situated at the true x-axis value.}
\label{mockscatter}
\end{figure*}

\begin{figure*}
  \centering
  \begin{tabular}{cc}
    \includegraphics[scale=0.65]{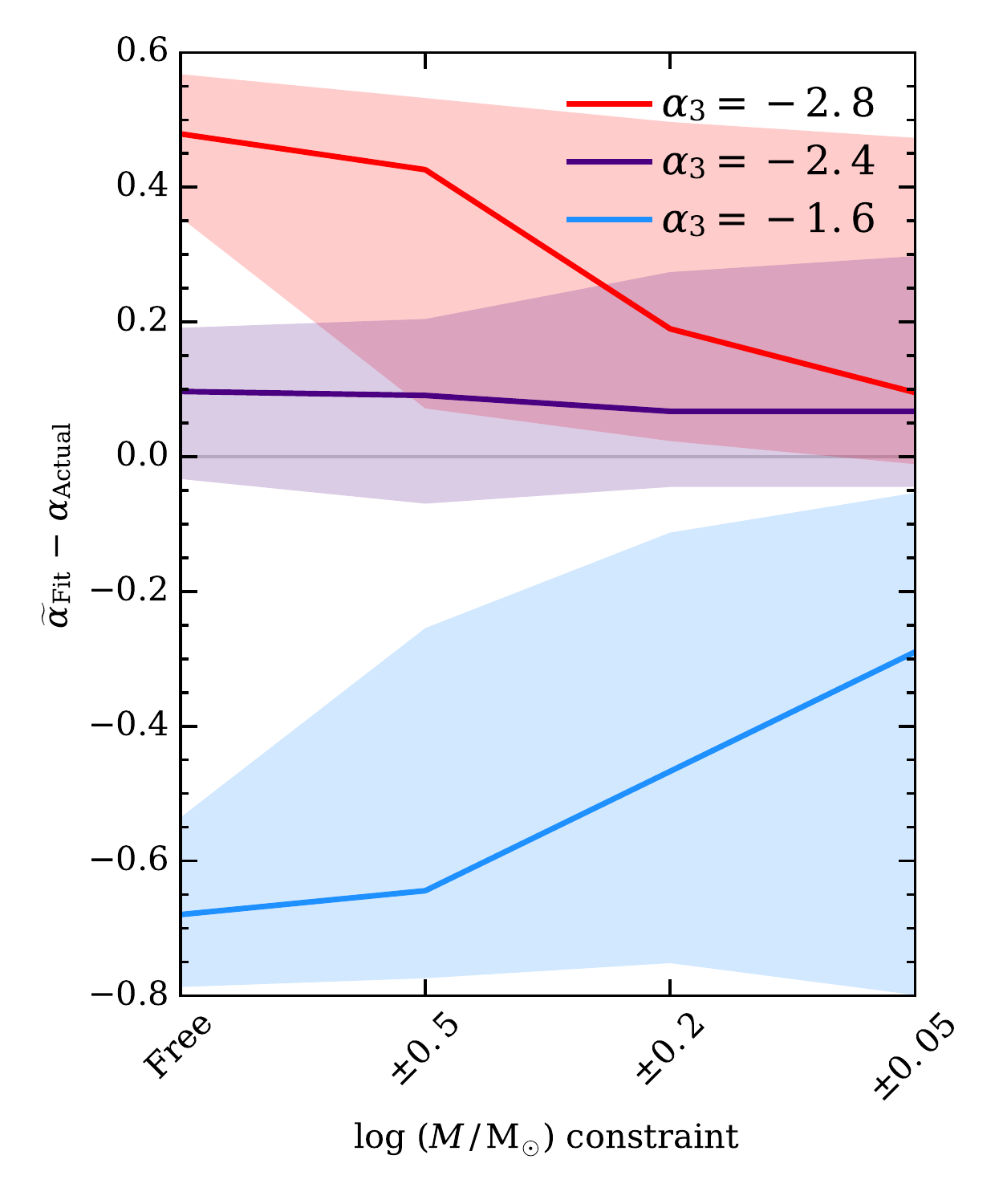} &
    \includegraphics[scale=0.65]{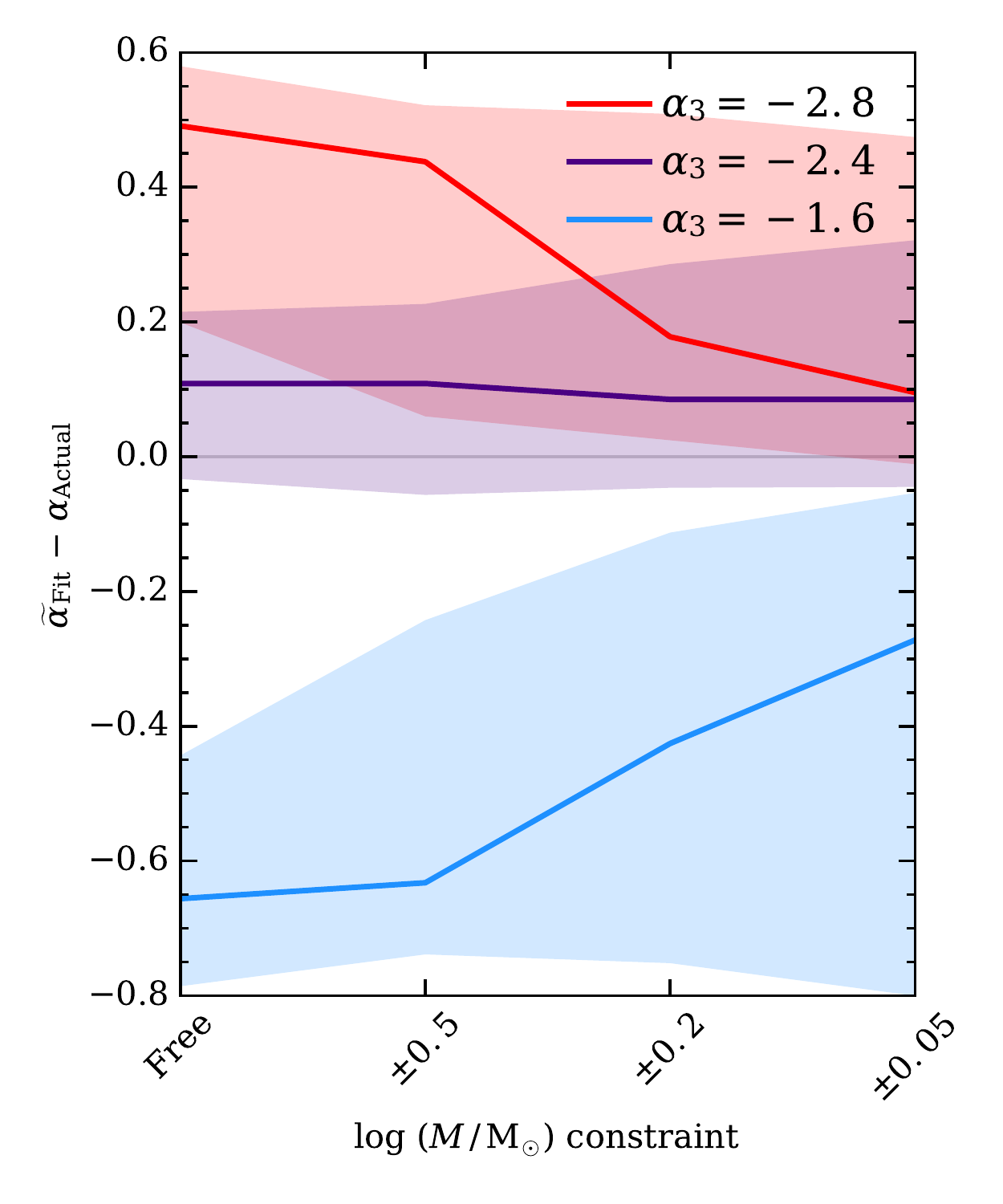}\\
    (a) Broad-band filters only & (b) Broad-band + \ha \\[6pt]
  \end{tabular}
\caption{Comparison between the IMF slopes $\alpha_3$ measured from the inferred posterior PDF (taking the median) and
  the true value as a function of width of a prior on the mass.  Results from three IMF slopes are shown.
  The filled regions represent the 10$^{\rm th}$ and 90$^{\rm th}$ percentile range. 
  We omit clusters with a target mass of $100\,\msun$ due to the relative variation in the resultant generated masses.
  Panel (a) was produced using only the 5 broad-band filters, whereas (b) was produced using the additional \ha\ filter. We see some improvement for the case with no constraints, with marginal improvement for the tighter constraints.}
\label{concom}
\end{figure*}
\begin{figure}
\centerline{\includegraphics[scale=0.7]{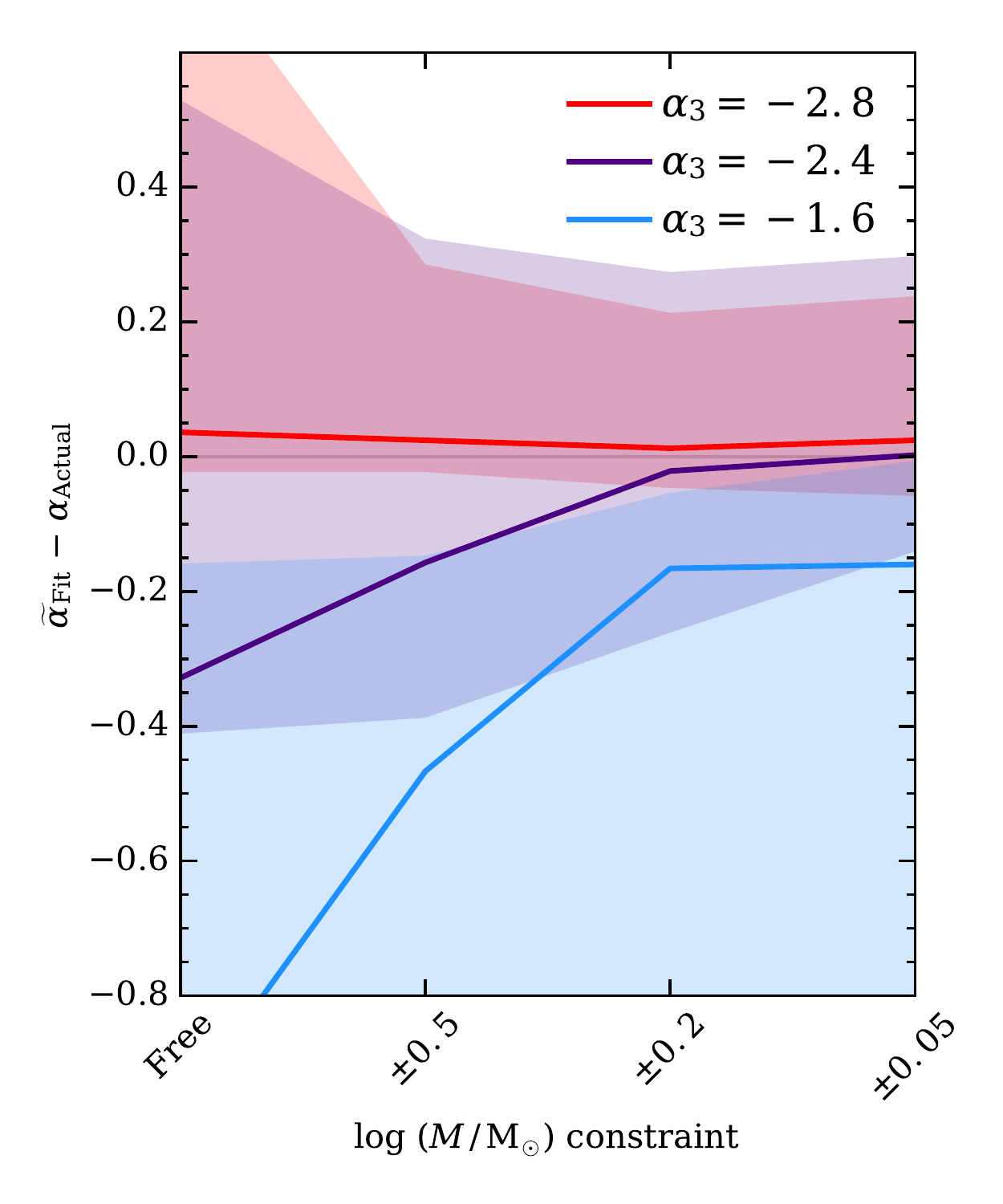}}
\centering
\caption{Comparison between the IMF slopes $\alpha_3$ measured from the inferred posterior PDF (taking the mode rather than the median) and  the true value as a function of width of a prior on the mass.  Results from three IMF slopes are shown.
  The filled regions represent the 10$^{\rm th}$ and 90$^{\rm th}$ percentile range. 
  We omit clusters with a target mass of $100\,\msun$ due to the relative variation in the resultant generated masses.}
\label{concompf}
\end{figure}

\section{Analysing the performance of the technique using mock clusters}\label{sec:predict}

Having applied our Bayesian formalism to the study of star clusters in NGC~628E, we have found that
a degeneracy between mass and the slope of the upper-end of the IMF prevents us from reliably constraining
the value of \imfs\, both when using broad-band photometry alone and when H$\alpha$ is included. We have also shown
how the choice of prior on the mass has a knock-on effect on the shape of the posterior PDF for \imfs.
Equivalently, the choice of a constant IMF may lead to systematic errors in the mass determination of more than
0.5 dex (depending on age).

In this section, we exploit mock clusters to take a more detailed look at our ability to
recover the posterior PDFs for physical parameters of clusters using our Bayesian formalism.
We also characterise the significance of the mass-\imfs\ degeneracy, providing forecasts on the
ability to constrain the IMF in future experiments when the cluster mass can be determined
via techniques that are independent of photometry. 

To achieve these goals, we generate a grid of 700 mock star clusters using \slug, with masses in the range
$\log (M/\rm{M}_{\odot})=2-6$ in steps of 1 dex, ages in the range $\log (T/\rm{yr})=6-9$ in steps of 0.5 dex,
and five choices of extinction, $A_V=0.1, 0.5, 1.0, 1.5, 2.0$ mag. For the IMF upper-end slope ($\alpha_3$), we select values
ranging from -2.8 to -1.6 in steps of 0.2. These parameters, together with the other \slug\ parameters that define
the SPS calculation, are listed in Table~\ref{tab:mockparam}.

We also generate a second grid of 175 clusters with a constant Kroupa IMF ($\alpha_3=-2.3$) spanning the same ranges of mass, age, and extinction. The parameters for this grid are also listed in Table~\ref{tab:mockparam}. 
Due to the stochastic nature of the \slug\ code, clusters simulated with a given choice of parameters do not necessarily have identical properties. For this reason, we generate 3 realisations of each cluster across both of our grids to help prevent the inclusion of rare outliers in our analysis. 
Using the variable library generated for the analysis of NGC~628E in Section~\ref{sec:libbuild}, we apply \clus\ to our grid of mock clusters with a variable IMF. Likewise, we apply \clus\ to the fixed IMF grid using the fiducial library of \cite{Krumholz2015-SLUGandLEGUS} as a training set.
We use flat priors during this analysis as the clusters are drawn from a flat distribution of physical parameters, without following, for example, a cluster mass function.

At first, we examine the precision with which \clus\ can recover the physical parameters of the clusters. This is achieved by
studying the intrinsic width of the posterior PDFs recovered by our Bayesian analysis.
A sharp PDF not only means that physical parameters can be recovered to high precision, but also
ensures that the global properties of a sample (e.g. via the cumulative PDFs) reflect the
intrinsic scatter of the population. Conversely, very broad PDFs reflect not only a low precision
in recovering parameters, but also the inability to characterise the underlying scatter
of physical parameters within the cluster population. 
In practice, we stack both the one-dimensional and two-dimensional posterior PDFs returned by \clus\ for all 700 mock clusters
with a variable IMF. The PDFs are median-centred before stacking, and to select 
 a representative group of objects, we choose for each cluster the realisation 
 corresponding to the the second-best `fit' in mass (i.e. second-smallest residual).
 These stacks are shown in Figure~\ref{2dstacks} and Figure~\ref{2dstacksYMnHa}.

 Inspecting the 1D posterior PDFs recovered from broad-band photometry alone (Figure~\ref{2dstacks}),
 we see that the PDFs for the mass, age, and extinction have a well-defined peak. Conversely, the posterior PDF for \imfs\ is broad
 and flat, confirming that broad-band photometry alone has weak constraining power on the upper end of the
 IMF. Additional insight is offered by the analysis of the joint PDFs. The well-known degeneracy
 in age-extinction is recovered by our analysis, as is the degeneracy between mass and \imfs\ discussed
 in the previous sections. Further, the contours on the joint PDFs of age and extinction with \imfs\
 indicate that $T$ and $A_{\rm V}$ are largely insensitive to variations in the upper end slope of the IMF.

 Restricting our analysis to a subset of mock clusters with young ages ($T \le 10^{6.5}~\rm yr$,  Figure
 \ref{2dstacksYMnHa}), we see that the performance of our Bayesian technique improves, with the
 posterior PDFs of all the parameters becoming narrower. Qualitatively, however, the majority
 of the trends we see in Figure~\ref{2dstacks} remain. By comparing the posterior PDFs computed
 with and without the inclusion of H$\alpha$ (right and left panels, respectively), we
 see that H$\alpha$ is critical in improving the age determination,
 breaking the age-extinction degeneracy. The broadness in the age PDF seen in the left panel of Figure~\ref{2dstacksYMnHa} is reduced greatly following the inclusion of \ha, which is seen in the narrowing of the age PDF in the panel on the right. In turn, a better constraint in age (and extinction) is reflected in a narrower mass PDF, although it remains broad due to the mass-IMF degeneracy.
 Indeed, with refined PDFs on age and extinction, an evident ``banana'' shape emerges for the joint PDF for $M-\alpha_3$,
 which clearly highlights the degeneracy between mass and IMF that has shaped much of the discussion in this work. 
 By means of our Bayesian analysis, we can also accurately quantify this degeneracy (see below), concluding that
 a variation in $\Delta \alpha_3 \sim \pm 0.5$ maps into a mass variation of $\Delta \log (M/\msun) \sim \pm 1$ dex.

 Next, we consider how accurate our Bayesian procedure is in recovering the underlying physical
 parameters for each cluster. This is achieved by comparing the medians of the inferred posterior PDFs
 to the true value used when computing the mocks. As was done for the stacks, in an effort to apply our analysis to a
 representative sample of clusters, we consider for each set of parameters the realisation that yields the second-best ``fit'' to the
 mass. These results are presented in Figure~\ref{mockscatter}.  Looking at all four physical parameters,
 we see that the medians of the inferred PDFs are a trustworthy indicator for the true value of the mass, age, and extinction
 when considering the average population. For the case of constant IMF (purple diamonds), the scatter
 between individual clusters is modest, of the order of $\sim 0.2-0.3$ dex (mag). 
 Conversely,  the addition of a variable IMF parameter greatly increases the scatter of residuals on a cluster by cluster basis
 when compared to the grid of Kroupa IMF clusters. This increase in scatter is especially noticeable in mass, and originates
 from the $M-\alpha_3$ degeneracy described above. As the mass of the cluster and its IMF slope are highly dependent on each other, the wide
 scatter in mass is further reflected in the inaccurate values inferred for \imfs. Indeed, given the
 flat and broad posterior PDFs for the IMF slope, the medians of \imfs\ have no physical meaning and
 simply reflect the median of the prior on \imfs\ (in this case a flat distribution between $\alpha_3 = (-3.0,-1.5)$).

 Given this result, it is clear that obtaining refined constraints on the upper-end slope of the IMF requires a different approach. 
 It is readily apparent that the ability to apply a prior to the cluster mass that is independent of the broad-band photometry may prove very useful in analysis such as this. Independent constraints on the masses of clusters that may then be used as priors may come, for instance, from dynamical measurements \citep[e.g.][]{Ho1996,Gerhard2000}. The era of $30\,\rm{m}$ telescopes will make such measurements more feasible.

 To illustrate this point in more quantitative
 terms, we present in Figure~\ref{concom} a study of the accuracy with which \imfs\ can be
 recovered as a function of the width of a prior on the mass. 
 Results of three choices of IMF are presented, when leaving the mass unconstrained,
 and when applying a top-hat prior with width 0.5, 0.2, and 0.05 dex about the target cluster mass, for both broad-band filters only and broad-band filters combined with \ha.
 This analysis shows that, as expected, the medians of the PDFs on \imfs\ converge to the true value as the precision with which
 the mass is constrained increases. Due to the difference between logarithmic intervals in mass and the linear scale in the IMF upper-end
 exponent, there is a weak scaling of \imfs\ with the prior width, with the underlying values being recovered to within
 $\Delta \alpha_3 \sim \pm 0.2$ when the mass is constrained to better than $10\%$. However,
 more realistic constraints on the mass (to within a factor of 2), for instance via dynamical measurements,
 appear to be useful to rule out extreme variations in the IMF upper-end slope through multiwavelength broad-band 
 photometry, given the assumptions of our model. 
 The inclusion of \ha\ results in a wider scatter towards zero for the unconstrained case. This advantage becomes much less evident as the mass is constrained however. Finally, in Figure~\ref{concompf}, the mode rather than the median of the \imfs\ posterior PDF is used to represent the recovered \imfs. Here we see a much faster convergence than we see in Figure~\ref{concom}. The scatter also reduces rapidly as we
 tighten the constraints on the mass, although this improvement is negligible in the $\alpha_3=-1.6$ case, where the
 scatter remains very large.

%==============================================================================

\section{Summary and conclusions}\label{sec:conclusions}

In this paper, we have developed a Bayesian formalism to study the properties of the IMF in star clusters with
multiwavelength broad-band photometry combined with \slug\ stochastic stellar population synthesis models.
To investigate variation in the IMF, we first extended the capabilities of \slug,
implementing a flexible algorithm by which the PDFs defining physical functions such as the IMF
can be set to variable mode. With this variable mode, it is now possible to perform simulations of clusters with
an IMF for which specific parameters can be
varied continuously across a desired range.

Exploiting this variable mode, which is now part of the main distribution of \slug, 
we constructed a library of 20 million star clusters, simulated with a  Kroupa-like IMF for which
the slope of the high-mass end (\imfs) was randomly chosen for each cluster, evenly selecting between $-3.0$ and $-1.5$.
With this library, we then exploited the \clus\ Bayesian analysis framework
described in \cite{Krumholz2015-SLUGandLEGUS} to compute the posterior PDF of \imfs\ for clusters with observed
multiwavelength broad-band photometry.

As a proof of concept, we applied this formalism to a catalogue of 225 star clusters from the local galaxy NGC~628,
for which broad-band and H$\alpha$ photometry is available from the LEGUS \citep{Calzetti2015-LEGUSi}
and H$\alpha$-LEGUS (Chandar et al. in preparation) programmes. 
After assuming a physically motivated prior for the mass and age of star clusters, 
we found that our formalism recovers the PDFs of the core physical parameters of mass, age, and extinction
for the entire sample, consistent with those resulting from the analysis of the same clusters with a constant IMF \citep{Krumholz2015-SLUGandLEGUS}.
Conversely, we found very broad and flat PDFs for \imfs, indicating that broad-band photometry alone is
unable to tightly constrain the slope of the upper-end of the IMF. This result is primarily driven by
a noticeable degeneracy between mass and \imfs.

Due to the interlink between mass and IMF slope, we found that the posterior PDF for \imfs\ is quite sensitive
to the choice of priors. Compared to the assumption of a flat prior in mass, a physically motivated
choice derived from the power-law nature of the cluster mass function leads to posterior PDFs
for the mass that are systematically skewed towards lower cluster masses and with a preference
for more shallow IMF slopes, in excess of the canonical Kroupa value of \imfs$=-2.3$. However, the fact that
the posterior PDF is very broad, and that it is sensitive to the choice of prior cautions against
far-reaching conclusions on the nature of the IMF with present data. Conversely, applying a prior to the cluster ages alone has a negligible effect on our analysis.

Adding \ha\ fluxes to our analysis does not significantly narrow the posterior PDF for \imfs, although
it reduces the degeneracy between age and extinction. As a consequence,
the posterior PDFs for age and extinction, and to some extent mass, sharpen. This effect
is particularly noticeable for the subsets of young star clusters, with ages $T \le 10^{6.5}~\rm yr$.

We also quantified the accuracy and precision with which our procedure is able to recover the input physical parameters for a grid of mock clusters, simulated across a wide range of physical parameters. We find the posterior PDFs for mass, age, and extinction have, on average, well defined medians. The width of the peaks in the PDFs are particularly narrow when considering a subset of young clusters, and when including \ha\ in the analysis. We find that we accurately recover the input parameters of our mock clusters, on average, with the medians of the mass, age, and extinction agreeing well with the input parameters. However, we still experience difficulty in recovering our input \imfs, finding a very broad posterior PDF across many mock clusters.

Finally, we have presented future avenues for improvement in our ability to constrain the slope of the upper end of the IMF via broad-band photometry. One possible path, rather than performing the analysis of the photometry on a cluster-by-cluster basis as we do in this study, is to perform a joint analysis of the star clusters with an underlying mass function that is common to all. This ensemble method could lessen the impact of the mass prior on our results.
  Furthermore, the ability to break the degeneracy between mass and \imfs\ via priors on the mass that are independent of the photometry (such as dynamical cluster mass measurements) is expected to provide us with the best way forward in our attempts to constrain the IMF slope by photometry. Indeed, we have explicitly shown that, while mass constraints of 10\% are required to reliably recover \imfs, priors that restrict the allowed mass range to within a factor of two are sufficient to rule out extreme variations in the allowed range of \imfs.

In conclusion, while our ability to recover the slope of the upper-end of the IMF
via broad-band photometry is limited at present time, there are good prospects
to further develop the formalism we have presented here to obtain improved
constraints on \imfs\ with current data, or via future spectroscopic
observations in the era of 30\,m telescopes. 

%==============================================================================
\section*{Acknowledgements}

We would like to thank R. Smith for informative discussions, and the anonymous referee for their helpful suggestions.

G.A. acknowledges support from the Science and Technology Facilities Council (ST/L00075X/1 and ST/M503472/1).
M.F. acknowledges support by the Science and Technology Facilities Council [grant number ST/L00075X/1].
D.A.G. kindly acknowledges financial support by the German Research Foundation (DFG) through program GO\,1659/3-2.

This work used the DiRAC Data Centric system at Durham University, 
operated by the Institute for Computational Cosmology on behalf of the 
STFC DiRAC HPC Facility (www.dirac.ac.uk). This equipment was funded by 
BIS National E-infrastructure capital grant ST/K00042X/1, STFC capital 
grants ST/H008519/1 and ST/K00087X/1, STFC DiRAC Operations grant 
ST/K003267/1 and Durham University. DiRAC is part of the National 
E-Infrastructure.

Based on observations obtained with the NASA/ESA Hubble Space Telescope at the Space Telescope Science Institute, which is operated by the Association of Universities for Research in Astronomy under NASA Contract NAS 5-26555. 

This research has made use of the NASA/IPAC Extragalactic Database (NED), which is operated by the Jet Propulsion Laboratory, California Institute of Technology, under contract with the National Aeronautics and Space Administration.

For access to the data and codes used in this work, please contact the authors or visit 
http://www.slugsps.com.

\bibliographystyle{mnras}
\bibliography{paper2bib}
%==============================================================================

\appendix
\section{Implementing a continuously variable IMF in SLUG} \label{sec:variable}

As described in Section~\ref{sec:introvariable}, \slug\ is a stochastic stellar population synthesis code which we make use of in our study of IMF variations. This requires a large library of simulated clusters with a wide variety of IMF shapes.

While in its previous version \slug\ already handled arbitrary IMF shapes,
the parameters specifying the functional form of the IMF were fixed by an input parameter file,
which prevented the user from running large simulations with a continuous distribution of clusters
with respect to the parameters defining the shape of the IMF. As an example, consider the Kroupa
IMF constructed with three power-law segments, having slopes $\alpha_1 = -0.3$, $\alpha_2 = -1.3$,
and $\alpha_3 = -2.3$. If we wished to investigate variation in $\alpha_3$, it would have been
necessary to run many libraries of \slug\ simulations, each configured with a different input value
of $\alpha_3$. This is clearly not desirable, especially when dealing with simulations of millions of clusters.  To circumvent this limitation, we develop a new feature of \slug\
which allows users to vary the parameters that control the IMF continuously, in a similar way to what is done
with the other physical parameters (e.g. mass, age, and extinction).

To this end, we implement an extension of \slug's PDF capabilities, introducing nested PDFs. In \slug, PDFs are generated and handled in a very general way \citep{daSilva2012-SLUGi,daSilva2014-SLUGii,Krumholz2015-SLUGiii}. Each PDF is described by an arbitrary
number of segments, with each segment having a functional form of choice,
such as a log-normal distribution or a power-law distribution \citep[a full list of functional forms is given by][]{Krumholz2015-SLUGiii}. In the code, these PDFs are used to represent the IMF,
the CMF, Star Formation History (SFH), distribution of extinctions ($A_{\rm{V}}$), and the Cluster Lifetime Function (CLF).

Our extension is implemented by enabling a hierarchy of PDFs, by which a ``master'' PDF can control, at run time, the behaviour of ``slave'' PDFs. With this new feature, users can define the functional form of
the IMF (the ``slave'' PDF) by means of an arbitrary number of segments as before. However, one or more parameters defining these segments can now be set to a variable mode. The user further specifies
the form of one ``master'' PDF for each variable parameter, from which the numerical values for
the variable parameter are drawn during the simulation. 
Specifically, at run time, \slug\ draws new realisations for each of the variable parameters
from the ``master'' PDFs and reinitialises the ``slave'' PDFs (the IMF in our case)
for each cluster accordingly. The end result of this implementation when applied to the IMF case
is that we can easily construct libraries of simulations with continuously varying IMFs, using only an input parameter file. 

Operationally, the use of this variable feature merely requires users to set
the correct flag when defining the PDF segments in the \slug\ IMF definition file, and then to create a companion file to define the PDF from which the parameter is drawn. Detailed instructions for the use of this new feature are available in the latest release of the \slug\ user manual\footnote{Available at http://slug2.readthedocs.io}, along with examples. The source code featuring this extension is publicly available from the \slug\ repository, and it is now part of the main
\slug\ release\footnote{Available at http://www.slugsps.com}. 

\begin{figure}
\centerline{\includegraphics[scale=0.7]{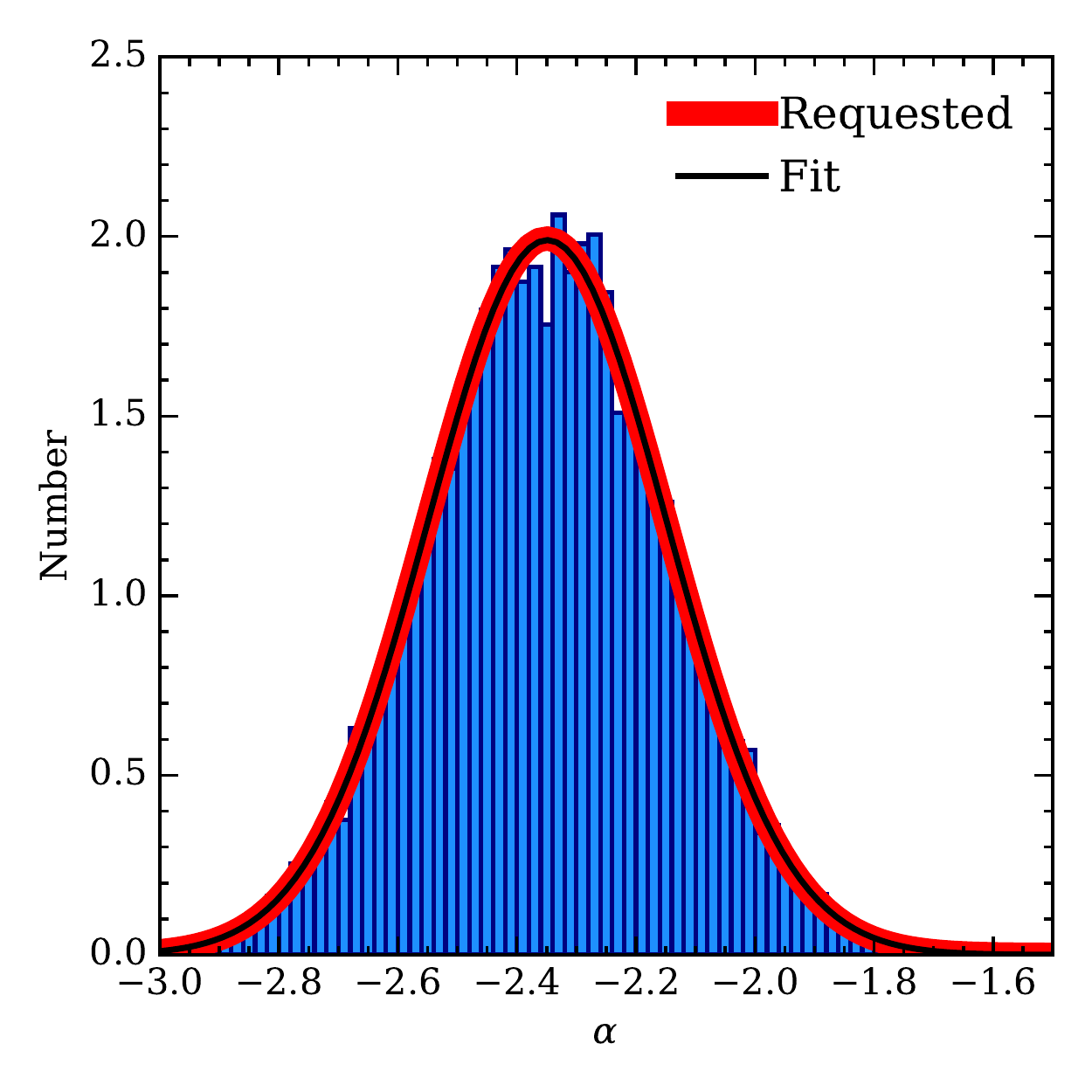}}
\centering
\caption{Values of the power law index $\alpha$ of a variable Salpeter-like IMF (blue histogram), obtained from $10^4$  cluster realisations generated using the newly-developed variable mode in \slug. The requested normal distribution for $\alpha$ has a mean of -2.35 and a dispersion of 0.2, and is plotted in red. The best fit normal distribution is displayed in black, and closely agrees with the input PDF.}
\label{normdraw}
\end{figure}

\begin{figure}
\centerline{\includegraphics[scale=0.7]{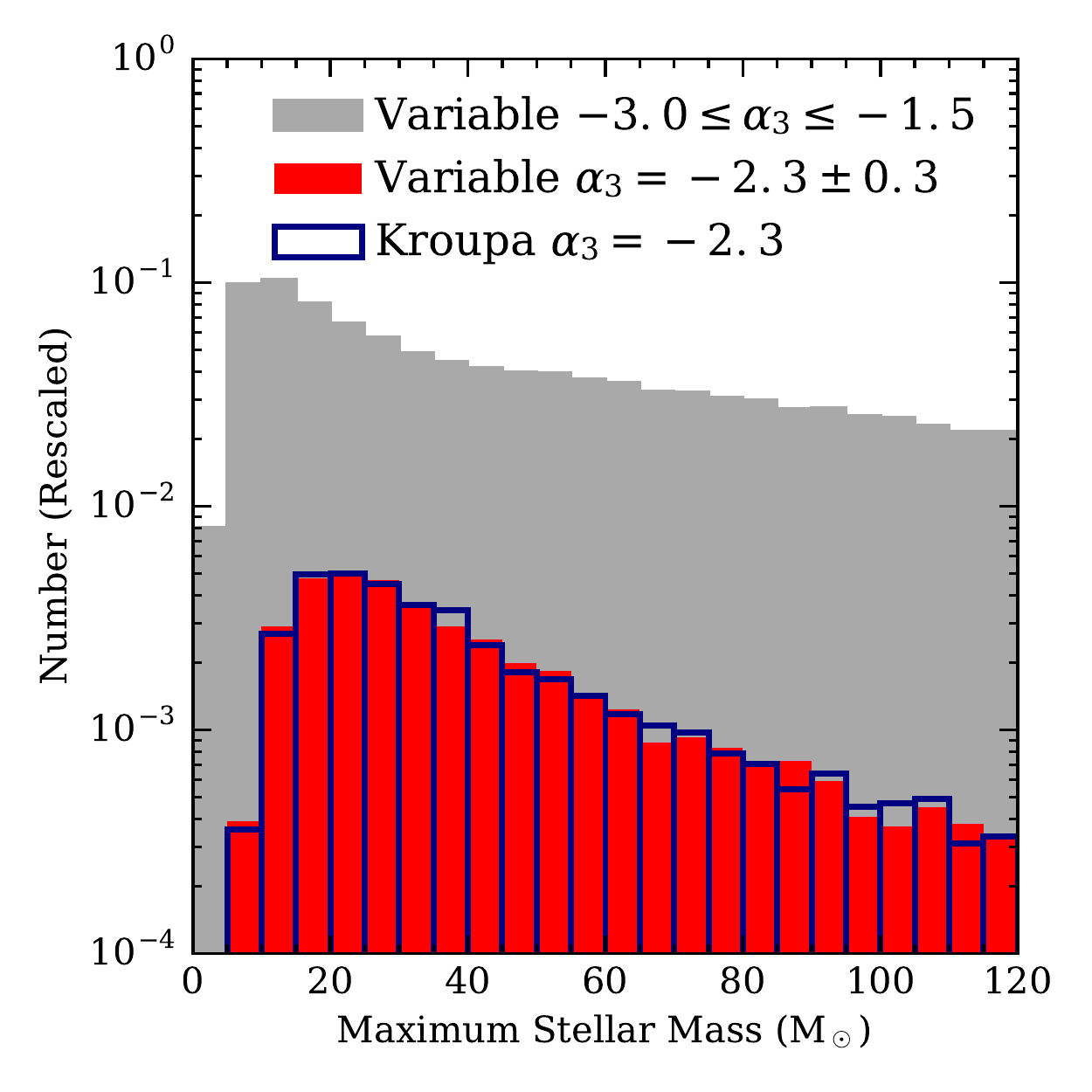}}
\centering
\caption{Comparison of the maximum stellar mass for two sets of realisations of a
  $500\,\rm{M}_{\odot}$ cluster at $10^6\,\rm{yr}$. In the first set (blue histogram),
  the stars are drawn from a canonical Kroupa IMF in 5000 realisations. In the second set, we
  generate $10^5$ realisations using the variable IMF mode, with the upper-end slope $\alpha_3$
  drawn from a flat distribution between $-3.0$ and $-1.5$. The subset of realisations with a
  value of $\alpha_3$ within $\pm0.3$ of the Kroupa value is shown in red, while the full range
  is shown in grey. The histograms are rescaled such that there are an equal number of clusters in the bins for the Kroupa and constrained variable cases, such that their size is relative to the complete variable set.}
\label{varycom}
\end{figure}

Here we provide examples of the capabilities of this variable mode,
together with tests to validate the new version of the code.  As a first example,
we consider the simple case of a Salpeter-like IMF \citep{Salpeter1955-IMF},
which is defined by a single power-law segment with slope $\alpha$.
In this run, we set $\alpha$ to variable mode, further providing a Gaussian PDF with
mean $\alpha=-2.35$ and dispersion of $0.2$ as the ``master'' PDF. The values of $\alpha$ drawn in
$10^4$ cluster realisations of a single \slug\ run are shown to follow closely the requested normal distribution in Figure~\ref{normdraw}. 

Next, we demonstrate the use of the variable PDFs for the case of an IMF with multiple segments.
Due to our flexible implementation, we can choose to vary only one of the parameters of this IMF.
In this demonstration, we vary only the high-mass slope (\imfs) of a Kroupa-like IMF, with its slope generated by evenly selecting between $-3$ and $-1.5$ for each cluster realisation. We generate $10^5$ realisations of a $500\,\msun$ star cluster, and we plot histograms of the most massive star in each
simulated cluster in Figure~\ref{varycom}. Given enough realisations, we can extract from the
resulting library of \slug\ models a subset of simulations with $\alpha_3 \approx -2.3$, which we
then compare to simulations that have been run in the non-variable mode using a canonical
Kroupa IMF. We find excellent agreement when comparing the shapes of the two distributions, which validates
our new implementation.  Figure~\ref{varycom} also shows that, as expected, the inclusion of 
shallow IMF slopes in the library, with \imfs\ up to values of $-1.5$,  increases the probability of drawing
massive stars compared to a normal Kroupa IMF, skewing the maximum stellar mass distribution to
high values. 

In summary, our extension to the way \slug\ handles PDFs enables users to create large libraries
of clusters simulated using a continuous distribution of parameters defining the functional form of the IMF. These libraries can then be used jointly with
Bayesian analysis to derive posterior PDFs for the IMF parameters via comparisons with
broad-band photometry. 
%==============================================================================
\label{lastpage}

\end{document}

%% file: table_reflib.tex
\begin{table*}
\begin{center}
  \begin{tabular}{ l | c | c | c | c | c | c | c | c | c | c}
    \hline
    Name & Tracks & Base IMF & Z & Extinction$^{\rm{a}}$ & $\phi$$^{\rm{b}}$ & $\log M$ & $\log T$ & $A_V$ & \# Realisations & Variable Parameters \\ 
     & & & & & & (M$_{\odot}$) & (yr) & (mag) &  &  \\ \hline \vspace{0.1cm}
pad\_020\_vkroupa\_MW$^{\rm{c}}$ & Padova AGB  & Kroupa & 0.020 & MW & 0.5 & 2--8 & 5--10.18 & 0--3 & $2\times 10^8 $ & $-3.0 \le \alpha_3 \le {-1.5}$ \\
pad\_020\_kroupa\_MW$^{\rm{d}}$ & Padova AGB & Kroupa & 0.020 & MW & 0.5 & 2--8 & 5--10.18 & 0--3 & $1\times 10^7 $ & $-$ \\ 
    \hline
\multicolumn{11}{l}{ $^{\rm{a}}$ MW = Milky Way extinction curve.} \\
\multicolumn{11}{l}{ $^{\rm{b}}$ $\phi$ is the fraction of ionising photons that produce nebular emission within the aperture, combining the effects of a covering fraction less than unity} \\
\multicolumn{11}{l}{~~ and some portion of the ionising photons being absorbed directly by dust.} \\ 
\multicolumn{11}{l}{ $^{\rm{c}}$ Variable IMF model.}\\
\multicolumn{11}{l}{ $^{\rm{d}}$ Constant IMF model \citep{Krumholz2015-SLUGandLEGUS}.}\\
  \end{tabular}
\end{center}
\caption{Table of the parameters used in generating the two reference libraries of \slug\ star clusters used in this work.}
\label{tab:simparam}
\end{table*}

%% file: table_mocks.tex
\begin{table*}
\begin{center}
  \begin{tabular}{ l | c | c | c | c | c | c | c | c | c | c}
    \hline
    Name & Tracks & Base IMF & Z & Extinction$^{\rm{a}}$ & $\phi$$^{\rm{b}}$ & $\log M $ & $\log T $ & $A_V$ & \#$^{\rm{e}}$  & \imfs \\ 
     & & & & & & (M$_{\odot}$) & (yr) & (mag) &  &  \\ \hline \vspace{0.1cm}
MocksRun1$^{\rm{c}}$ & Padova AGB  & Kroupa & 0.020 & MW & 0.5 & 2--6\,(1) & 6--9\,(0.5) & 0.1,0.5,1,1.5,2 & $ 3 $ & $-2.8,-1.6 (0.2)$\\
MocksRun2$^{\rm{d}}$ & Padova AGB & Kroupa & 0.020 & MW & 0.5 & 2--6\,(1) & 6--9\,(0.5) & 0.1,0.5,1,1.5,2 & $ 3 $ & $ -2.3$ \\ 
    \hline
\multicolumn{11}{l}{ $^{\rm{a}}$ MW = Milky Way extinction curve} \\
\multicolumn{11}{l}{ $^{\rm{b}}$ $\phi$ is the fraction of ionising photons that produce nebular emission within the aperture, combining the effects of a covering fraction less than 1} \\
\multicolumn{11}{l}{~~ and some portion of the ionising photons being absorbed directly by dust} \\ 
\multicolumn{11}{l}{ $^{\rm{c}}$ Variable IMF mocks}\\
\multicolumn{11}{l}{ $^{\rm{d}}$ Constant IMF mocks}\\
\multicolumn{11}{l}{ $^{\rm{e}}$ Number of realisations}\\
  \end{tabular}
\end{center}
\caption{Table of the parameters used in \slug\ for generating the two grids of mock star clusters used in this work. Within the parentheses we list the adopted step size.}
\label{tab:mockparam}
\end{table*}

%% file: paper2.bbl
\begin{thebibliography}{}
\makeatletter
\relax
\def\mn@urlcharsother{\let\do\@makeother \do\$\do\&\do\#\do\^\do\_\do\%\do\~}
\def\mn@doi{\begingroup\mn@urlcharsother \@ifnextchar [ {\mn@doi@}
  {\mn@doi@[]}}
\def\mn@doi@[#1]#2{\def\@tempa{#1}\ifx\@tempa\@empty \href
  {http://dx.doi.org/#2} {doi:#2}\else \href {http://dx.doi.org/#2} {#1}\fi
  \endgroup}
\def\mn@eprint#1#2{\mn@eprint@#1:#2::\@nil}
\def\mn@eprint@arXiv#1{\href {http://arxiv.org/abs/#1} {{\tt arXiv:#1}}}
\def\mn@eprint@dblp#1{\href {http://dblp.uni-trier.de/rec/bibtex/#1.xml}
  {dblp:#1}}
\def\mn@eprint@#1:#2:#3:#4\@nil{\def\@tempa {#1}\def\@tempb {#2}\def\@tempc
  {#3}\ifx \@tempc \@empty \let \@tempc \@tempb \let \@tempb \@tempa \fi \ifx
  \@tempb \@empty \def\@tempb {arXiv}\fi \@ifundefined
  {mn@eprint@\@tempb}{\@tempb:\@tempc}{\expandafter \expandafter \csname
  mn@eprint@\@tempb\endcsname \expandafter{\@tempc}}}

\bibitem[\protect\citeauthoryear{{Adams} \& {Fatuzzo}}{{Adams} \&
  {Fatuzzo}}{1996}]{Adams1996-IMFTheory}
{Adams} F.~C.,  {Fatuzzo} M.,  1996, \mn@doi [\apj] {10.1086/177318}, \href
  {http://adsabs.harvard.edu/abs/1996ApJ...464..256A} {464, 256}

\bibitem[\protect\citeauthoryear{{Andrews} et~al.,}{{Andrews}
  et~al.}{2013}]{Andrews2013-IMFinNGC4214}
{Andrews} J.~E.,  et~al., 2013, \mn@doi [\apj] {10.1088/0004-637X/767/1/51},
  \href {http://adsabs.harvard.edu/abs/2013ApJ...767...51A} {767, 51}

\bibitem[\protect\citeauthoryear{{Andrews} et~al.,}{{Andrews}
  et~al.}{2014}]{Andrews2014-IMFinM83}
{Andrews} J.~E.,  et~al., 2014, \mn@doi [\apj] {10.1088/0004-637X/793/1/4},
  \href {http://adsabs.harvard.edu/abs/2014ApJ...793....4A} {793, 4}

\bibitem[\protect\citeauthoryear{{Bastian}, {Covey}  \& {Meyer}}{{Bastian}
  et~al.}{2010}]{Bastian2010-IMF}
{Bastian} N.,  {Covey} K.~R.,   {Meyer} M.~R.,  2010, \mn@doi [\araa]
  {10.1146/annurev-astro-082708-101642}, \href
  {http://adsabs.harvard.edu/abs/2010ARA%26A..48..339B} {48, 339}

\bibitem[\protect\citeauthoryear{{Bell}, {McIntosh}, {Katz}  \&
  {Weinberg}}{{Bell} et~al.}{2003}]{Bell2003-Luminosity}
{Bell} E.~F.,  {McIntosh} D.~H.,  {Katz} N.,   {Weinberg} M.~D.,  2003, \mn@doi
  [\apjs] {10.1086/378847}, \href
  {http://adsabs.harvard.edu/abs/2003ApJS..149..289B} {149, 289}

\bibitem[\protect\citeauthoryear{{Bonnell}, {Clarke}  \& {Bate}}{{Bonnell}
  et~al.}{2006}]{Bonnell2006-IMFJeansMass}
{Bonnell} I.~A.,  {Clarke} C.~J.,   {Bate} M.~R.,  2006, \mn@doi [\mnras]
  {10.1111/j.1365-2966.2006.10214.x}, \href
  {http://adsabs.harvard.edu/abs/2006MNRAS.368.1296B} {368, 1296}

\bibitem[\protect\citeauthoryear{{Calzetti}, {Chandar}, {Lee}, {Elmegreen},
  {Kennicutt}  \& {Whitmore}}{{Calzetti} et~al.}{2010}]{Calzetti2010-IMF}
{Calzetti} D.,  {Chandar} R.,  {Lee} J.~C.,  {Elmegreen} B.~G.,  {Kennicutt}
  R.~C.,   {Whitmore} B.,  2010, \mn@doi [\apjl]
  {10.1088/2041-8205/719/2/L158}, \href
  {http://adsabs.harvard.edu/abs/2010ApJ...719L.158C} {719, L158}

\bibitem[\protect\citeauthoryear{{Calzetti} et~al.,}{{Calzetti}
  et~al.}{2015}]{Calzetti2015-LEGUSi}
{Calzetti} D.,  et~al., 2015, \mn@doi [\aj] {10.1088/0004-6256/149/2/51}, \href
  {http://adsabs.harvard.edu/abs/2015AJ....149...51C} {149, 51}

\bibitem[\protect\citeauthoryear{{Cappellari} et~al.,}{{Cappellari}
  et~al.}{2012}]{Cappellari2012-IMFinEarlyTypes}
{Cappellari} M.,  et~al., 2012, \mn@doi [\nat] {10.1038/nature10972}, \href
  {http://adsabs.harvard.edu/abs/2012Natur.484..485C} {484, 485}

\bibitem[\protect\citeauthoryear{{Cardelli}, {Clayton}  \& {Mathis}}{{Cardelli}
  et~al.}{1989}]{Cardelli1989-MWExt}
{Cardelli} J.~A.,  {Clayton} G.~C.,   {Mathis} J.~S.,  1989, \mn@doi [\apj]
  {10.1086/167900}, \href {http://adsabs.harvard.edu/abs/1989ApJ...345..245C}
  {345, 245}

\bibitem[\protect\citeauthoryear{{Cervi{\~n}o}, {Valls-Gabaud}, {Luridiana}  \&
  {Mas-Hesse}}{{Cervi{\~n}o} et~al.}{2002}]{Cervino2002-Sampling}
{Cervi{\~n}o} M.,  {Valls-Gabaud} D.,  {Luridiana} V.,   {Mas-Hesse} J.~M.,
  2002, \mn@doi [\aap] {10.1051/0004-6361:20011266}, \href
  {http://adsabs.harvard.edu/abs/2002A%26A...381...51C} {381, 51}

\bibitem[\protect\citeauthoryear{{Chabrier}}{{Chabrier}}{2003}]{Chabrier2003-IMF}
{Chabrier} G.,  2003, \mn@doi [\pasp] {10.1086/376392}, \href
  {http://adsabs.harvard.edu/abs/2003PASP..115..763C} {115, 763}

\bibitem[\protect\citeauthoryear{{Conroy} \& {van Dokkum}}{{Conroy} \& {van
  Dokkum}}{2012}]{Conroy2012-IMFinEarlyTypes}
{Conroy} C.,  {van Dokkum} P.~G.,  2012, \mn@doi [\apj]
  {10.1088/0004-637X/760/1/71}, \href
  {http://adsabs.harvard.edu/abs/2012ApJ...760...71C} {760, 71}

\bibitem[\protect\citeauthoryear{{Corbelli}, {Verley}, {Elmegreen}  \&
  {Giovanardi}}{{Corbelli} et~al.}{2009}]{corbelli09}
{Corbelli} E.,  {Verley} S.,  {Elmegreen} B.~G.,   {Giovanardi} C.,  2009,
  \mn@doi [\aap] {10.1051/0004-6361:200811086}, \href
  {http://adsabs.harvard.edu/abs/2009A%26A...495..479C} {495, 479}

\bibitem[\protect\citeauthoryear{{Dries}, {Trager}  \& {Koopmans}}{{Dries}
  et~al.}{2016}]{Dries2016-BayesianIMF}
{Dries} M.,  {Trager} S.~C.,   {Koopmans} L.~V.~E.,  2016, \mn@doi [\mnras]
  {10.1093/mnras/stw2049}, \href
  {http://adsabs.harvard.edu/abs/2016MNRAS.tmp.1169D} {}

\bibitem[\protect\citeauthoryear{{Ekstr{\"o}m} et~al.,}{{Ekstr{\"o}m}
  et~al.}{2012}]{genevarot}
{Ekstr{\"o}m} S.,  et~al., 2012, \mn@doi [\aap] {10.1051/0004-6361/201117751},
  \href {http://adsabs.harvard.edu/abs/2012A%26A...537A.146E} {537, A146}

\bibitem[\protect\citeauthoryear{{Eldridge} \& {Stanway}}{{Eldridge} \&
  {Stanway}}{2009}]{Eldridge2009-Binaries}
{Eldridge} J.~J.,  {Stanway} E.~R.,  2009, \mn@doi [\mnras]
  {10.1111/j.1365-2966.2009.15514.x}, \href
  {http://adsabs.harvard.edu/abs/2009MNRAS.400.1019E} {400, 1019}

\bibitem[\protect\citeauthoryear{{Elmegreen}}{{Elmegreen}}{2002}]{Elmegreen2002-StarFormation}
{Elmegreen} B.~G.,  2002, \mn@doi [\apj] {10.1086/342177}, \href
  {http://adsabs.harvard.edu/abs/2002ApJ...577..206E} {577, 206}

\bibitem[\protect\citeauthoryear{{Fumagalli}, {da Silva}  \&
  {Krumholz}}{{Fumagalli} et~al.}{2011}]{Fumagalli2011-SLUGLetter}
{Fumagalli} M.,  {da Silva} R.~L.,   {Krumholz} M.~R.,  2011, \mn@doi [\apjl]
  {10.1088/2041-8205/741/2/L26}, \href
  {http://adsabs.harvard.edu/abs/2011ApJ...741L..26F} {741, L26}

\bibitem[\protect\citeauthoryear{{Gerhard}}{{Gerhard}}{2000}]{Gerhard2000}
{Gerhard} O.,  2000, in {Lan{\c c}on} A.,  {Boily} C.~M.,  eds,  Astronomical
  Society of the Pacific Conference Series Vol. 211, Massive Stellar Clusters.
  p.~12 (\mn@eprint {} {astro-ph/0007258})

\bibitem[\protect\citeauthoryear{{Girardi}, {Bressan}, {Bertelli}  \&
  {Chiosi}}{{Girardi} et~al.}{2000}]{Girardi2000-Padova}
{Girardi} L.,  {Bressan} A.,  {Bertelli} G.,   {Chiosi} C.,  2000, \mn@doi
  [\aaps] {10.1051/aas:2000126}, \href
  {http://adsabs.harvard.edu/abs/2000A%26AS..141..371G} {141, 371}

\bibitem[\protect\citeauthoryear{{Grasha} et~al.,}{{Grasha}
  et~al.}{2015}]{Grasha2015}
{Grasha} K.,  et~al., 2015, \mn@doi [\apj] {10.1088/0004-637X/815/2/93}, \href
  {http://adsabs.harvard.edu/abs/2015ApJ...815...93G} {815, 93}

\bibitem[\protect\citeauthoryear{{Gunawardhana} et~al.,}{{Gunawardhana}
  et~al.}{2011}]{Gunawardhana2011-GAMA}
{Gunawardhana} M.~L.~P.,  et~al., 2011, \mn@doi [\mnras]
  {10.1111/j.1365-2966.2011.18800.x}, \href
  {http://adsabs.harvard.edu/abs/2011MNRAS.415.1647G} {415, 1647}

\bibitem[\protect\citeauthoryear{{Gvaramadze}, {Weidner}, {Kroupa}  \&
  {Pflamm-Altenburg}}{{Gvaramadze} et~al.}{2012}]{Gvaramadze2012-Runaways}
{Gvaramadze} V.~V.,  {Weidner} C.,  {Kroupa} P.,   {Pflamm-Altenburg} J.,
  2012, \mn@doi [\mnras] {10.1111/j.1365-2966.2012.21452.x}, \href
  {http://adsabs.harvard.edu/abs/2012MNRAS.424.3037G} {424, 3037}

\bibitem[\protect\citeauthoryear{{Haas} \& {Anders}}{{Haas} \&
  {Anders}}{2010}]{Haas2010-IMFVar}
{Haas} M.~R.,  {Anders} P.,  2010, \mn@doi [\aap]
  {10.1051/0004-6361/200912967}, \href
  {http://adsabs.harvard.edu/abs/2010A%26A...512A..79H} {512, A79}

\bibitem[\protect\citeauthoryear{{Ho} \& {Filippenko}}{{Ho} \&
  {Filippenko}}{1996}]{Ho1996}
{Ho} L.~C.,  {Filippenko} A.~V.,  1996, \mn@doi [\apjl] {10.1086/310181}, \href
  {http://adsabs.harvard.edu/abs/1996ApJ...466L..83H} {466, L83}

\bibitem[\protect\citeauthoryear{{Kroupa}}{{Kroupa}}{2001}]{Kroupa2001-IMF}
{Kroupa} P.,  2001, \mn@doi [\mnras] {10.1046/j.1365-8711.2001.04022.x}, \href
  {http://adsabs.harvard.edu/abs/2001MNRAS.322..231K} {322, 231}

\bibitem[\protect\citeauthoryear{{Kroupa} \& {Weidner}}{{Kroupa} \&
  {Weidner}}{2003}]{Kroupa2003-IGIMF}
{Kroupa} P.,  {Weidner} C.,  2003, \mn@doi [\apj] {10.1086/379105}, \href
  {http://adsabs.harvard.edu/abs/2003ApJ...598.1076K} {598, 1076}

\bibitem[\protect\citeauthoryear{{Kroupa}, {Weidner}, {Pflamm-Altenburg},
  {Thies}, {Dabringhausen}, {Marks}  \& {Maschberger}}{{Kroupa}
  et~al.}{2013}]{Kroupa2013-StellarSubStellar}
{Kroupa} P.,  {Weidner} C.,  {Pflamm-Altenburg} J.,  {Thies} I.,
  {Dabringhausen} J.,  {Marks} M.,   {Maschberger} T.,  2013, {The Stellar and
  Sub-Stellar Initial Mass Function of Simple and Composite Populations}.
p.~115, \mn@doi{10.1007/978-94-007-5612-0_4}

\bibitem[\protect\citeauthoryear{{Krumholz}}{{Krumholz}}{2011}]{Krumholz2011-FragIMF}
{Krumholz} M.~R.,  2011, in {Treyer} M.,  {Wyder} T.,  {Neill} J.,  {Seibert}
  M.,   {Lee} J.,  eds,  Astronomical Society of the Pacific Conference Series
  Vol. 440, UP2010: Have Observations Revealed a Variable Upper End of the
  Initial Mass Function?. p.~91

\bibitem[\protect\citeauthoryear{{Krumholz}, {Fumagalli}, {da Silva}, {Rendahl}
   \& {Parra}}{{Krumholz} et~al.}{2015a}]{Krumholz2015-SLUGiii}
{Krumholz} M.~R.,  {Fumagalli} M.,  {da Silva} R.~L.,  {Rendahl} T.,   {Parra}
  J.,  2015a, \mn@doi [\mnras] {10.1093/mnras/stv1374}, \href
  {http://adsabs.harvard.edu/abs/2015MNRAS.452.1447K} {452, 1447}

\bibitem[\protect\citeauthoryear{{Krumholz} et~al.,}{{Krumholz}
  et~al.}{2015b}]{Krumholz2015-SLUGandLEGUS}
{Krumholz} M.~R.,  et~al., 2015b, \mn@doi [\apj] {10.1088/0004-637X/812/2/147},
  \href {http://adsabs.harvard.edu/abs/2015ApJ...812..147K} {812, 147}

\bibitem[\protect\citeauthoryear{{Lamb}, {Oey}, {Werk}  \& {Ingleby}}{{Lamb}
  et~al.}{2010}]{Lamb2010-SolitaryOB}
{Lamb} J.~B.,  {Oey} M.~S.,  {Werk} J.~K.,   {Ingleby} L.~D.,  2010, \mn@doi
  [\apj] {10.1088/0004-637X/725/2/1886}, \href
  {http://adsabs.harvard.edu/abs/2010ApJ...725.1886L} {725, 1886}

\bibitem[\protect\citeauthoryear{{Lamb}, {Oey}, {Segura-Cox}, {Graus},
  {Kiminki}, {Golden-Marx}  \& {Parker}}{{Lamb} et~al.}{2016}]{Lamb2016-RIOTS}
{Lamb} J.~B.,  {Oey} M.~S.,  {Segura-Cox} D.~M.,  {Graus} A.~S.,  {Kiminki}
  D.~C.,  {Golden-Marx} J.~B.,   {Parker} J.~W.,  2016, \mn@doi [\apj]
  {10.3847/0004-637X/817/2/113}, \href
  {http://adsabs.harvard.edu/abs/2016ApJ...817..113L} {817, 113}

\bibitem[\protect\citeauthoryear{{Lee} et~al.,}{{Lee}
  et~al.}{2009}]{Lee2009-HaFUV}
{Lee} J.~C.,  et~al., 2009, \mn@doi [\apj] {10.1088/0004-637X/706/1/599}, \href
  {http://adsabs.harvard.edu/abs/2009ApJ...706..599L} {706, 599}

\bibitem[\protect\citeauthoryear{{Lee}, {Veilleux}, {McDonald}  \&
  {Hilbert}}{{Lee} et~al.}{2016}]{Lee2016-HaFUV}
{Lee} J.~C.,  {Veilleux} S.,  {McDonald} M.,   {Hilbert} B.,  2016, \mn@doi
  [\apj] {10.3847/0004-637X/817/2/177}, \href
  {http://adsabs.harvard.edu/abs/2016ApJ...817..177L} {817, 177}

\bibitem[\protect\citeauthoryear{{Leitherer} et~al.,}{{Leitherer}
  et~al.}{1999}]{Leitherer1999-STARBURST99}
{Leitherer} C.,  et~al., 1999, \mn@doi [\apjs] {10.1086/313233}, \href
  {http://adsabs.harvard.edu/abs/1999ApJS..123....3L} {123, 3}

\bibitem[\protect\citeauthoryear{{Leitherer}, {Ekstr{\"o}m}, {Meynet},
  {Schaerer}, {Agienko}  \& {Levesque}}{{Leitherer}
  et~al.}{2014}]{Leitherer2014-Rotation}
{Leitherer} C.,  {Ekstr{\"o}m} S.,  {Meynet} G.,  {Schaerer} D.,  {Agienko}
  K.~B.,   {Levesque} E.~M.,  2014, \mn@doi [\apjs]
  {10.1088/0067-0049/212/1/14}, \href
  {http://adsabs.harvard.edu/abs/2014ApJS..212...14L} {212, 14}

\bibitem[\protect\citeauthoryear{{Levesque}, {Leitherer}, {Ekstrom}, {Meynet}
  \& {Schaerer}}{{Levesque} et~al.}{2012}]{Levesque2012-Rotation}
{Levesque} E.~M.,  {Leitherer} C.,  {Ekstrom} S.,  {Meynet} G.,   {Schaerer}
  D.,  2012, \mn@doi [\apj] {10.1088/0004-637X/751/1/67}, \href
  {http://adsabs.harvard.edu/abs/2012ApJ...751...67L} {751, 67}

\bibitem[\protect\citeauthoryear{{Li} \& {Han}}{{Li} \&
  {Han}}{2008}]{Li2008-Binaries}
{Li} Z.,  {Han} Z.,  2008, \mn@doi [\apj] {10.1086/590228}, \href
  {http://adsabs.harvard.edu/abs/2008ApJ...685..225L} {685, 225}

\bibitem[\protect\citeauthoryear{{Lyubenova} et~al.,}{{Lyubenova}
  et~al.}{2016}]{Lyubenova2016-CALIFA}
{Lyubenova} M.,  et~al., 2016, \mn@doi [\mnras] {10.1093/mnras/stw2434}, \href
  {http://adsabs.harvard.edu/abs/2016MNRAS.463.3220L} {463, 3220}

\bibitem[\protect\citeauthoryear{{Massey}, {Johnson}  \&
  {Degioia-Eastwood}}{{Massey} et~al.}{1995}]{Massey1995-IMF}
{Massey} P.,  {Johnson} K.~E.,   {Degioia-Eastwood} K.,  1995, \mn@doi [\apj]
  {10.1086/176474}, \href {http://adsabs.harvard.edu/abs/1995ApJ...454..151M}
  {454, 151}

\bibitem[\protect\citeauthoryear{{Matteucci}}{{Matteucci}}{1994}]{Matteucci1994-ChemEv1}
{Matteucci} F.,  1994, \aap, \href
  {http://adsabs.harvard.edu/abs/1994A%26A...288...57M} {288, 57}

\bibitem[\protect\citeauthoryear{{Meurer} et~al.,}{{Meurer}
  et~al.}{2009}]{Meurer2009-IMF}
{Meurer} G.~R.,  et~al., 2009, \mn@doi [\apj] {10.1088/0004-637X/695/1/765},
  \href {http://adsabs.harvard.edu/abs/2009ApJ...695..765M} {695, 765}

\bibitem[\protect\citeauthoryear{{Oey}, {Lamb}, {Kushner}, {Pellegrini}  \&
  {Graus}}{{Oey} et~al.}{2013}]{Oey2013-SoloOBStars}
{Oey} M.~S.,  {Lamb} J.~B.,  {Kushner} C.~T.,  {Pellegrini} E.~W.,   {Graus}
  A.~S.,  2013, \mn@doi [\apj] {10.1088/0004-637X/768/1/66}, \href
  {http://adsabs.harvard.edu/abs/2013ApJ...768...66O} {768, 66}

\bibitem[\protect\citeauthoryear{{Popescu} \& {Hanson}}{{Popescu} \&
  {Hanson}}{2009}]{Popescu2009-MASSCLEAN1}
{Popescu} B.,  {Hanson} M.~M.,  2009, \mn@doi [\aj]
  {10.1088/0004-6256/138/6/1724}, \href
  {http://adsabs.harvard.edu/abs/2009AJ....138.1724P} {138, 1724}

\bibitem[\protect\citeauthoryear{{Salpeter}}{{Salpeter}}{1955}]{Salpeter1955-IMF}
{Salpeter} E.~E.,  1955, \mn@doi [\apj] {10.1086/145971}, \href
  {http://adsabs.harvard.edu/abs/1955ApJ...121..161S} {121, 161}

\bibitem[\protect\citeauthoryear{{Smith}}{{Smith}}{2014}]{Smith2014-IMFVar}
{Smith} R.~J.,  2014, \mn@doi [\mnras] {10.1093/mnrasl/slu082}, \href
  {http://adsabs.harvard.edu/abs/2014MNRAS.443L..69S} {443, L69}

\bibitem[\protect\citeauthoryear{{Stephens} et~al.,}{{Stephens}
  et~al.}{2017}]{Stephens2017-OBStars}
{Stephens} I.~W.,  et~al., 2017, \mn@doi [\apj] {10.3847/1538-4357/834/1/94},
  \href {http://adsabs.harvard.edu/abs/2017ApJ...834...94S} {834, 94}

\bibitem[\protect\citeauthoryear{{Thomas}}{{Thomas}}{1999}]{Thomas1999-ChemEv1}
{Thomas} D.,  1999, \mn@doi [\mnras] {10.1046/j.1365-8711.1999.02552.x}, \href
  {http://adsabs.harvard.edu/abs/1999MNRAS.306..655T} {306, 655}

\bibitem[\protect\citeauthoryear{{Vassiliadis} \& {Wood}}{{Vassiliadis} \&
  {Wood}}{1993}]{Vassiladis1993-PadovaAGB}
{Vassiliadis} E.,  {Wood} P.~R.,  1993, \mn@doi [\apj] {10.1086/173033}, \href
  {http://adsabs.harvard.edu/abs/1993ApJ...413..641V} {413, 641}

\bibitem[\protect\citeauthoryear{{V{\'a}zquez} \& {Leitherer}}{{V{\'a}zquez} \&
  {Leitherer}}{2005}]{Vazquez2005-Padova}
{V{\'a}zquez} G.~A.,  {Leitherer} C.,  2005, \mn@doi [\apj] {10.1086/427866},
  \href {http://adsabs.harvard.edu/abs/2005ApJ...621..695V} {621, 695}

\bibitem[\protect\citeauthoryear{{V{\'a}zquez}, {Leitherer}, {Schaerer},
  {Meynet}  \& {Maeder}}{{V{\'a}zquez} et~al.}{2007}]{Vasquez2007-Rotation}
{V{\'a}zquez} G.~A.,  {Leitherer} C.,  {Schaerer} D.,  {Meynet} G.,   {Maeder}
  A.,  2007, \mn@doi [\apj] {10.1086/518589}, \href
  {http://adsabs.harvard.edu/abs/2007ApJ...663..995V} {663, 995}

\bibitem[\protect\citeauthoryear{{Weidner}, {Pflamm-Altenburg}  \&
  {Kroupa}}{{Weidner} et~al.}{2011}]{Weidner2011-IGIMF}
{Weidner} C.,  {Pflamm-Altenburg} J.,   {Kroupa} P.,  2011, in {Treyer} M.,
  {Wyder} T.,  {Neill} J.,  {Seibert} M.,   {Lee} J.,  eds,  Astronomical
  Society of the Pacific Conference Series Vol. 440, UP2010: Have Observations
  Revealed a Variable Upper End of the Initial Mass Function?. p.~19
  (\mn@eprint {arXiv} {1011.1905})

\bibitem[\protect\citeauthoryear{{Weisz}, {Fouesneau}, {Hogg}, {Rix},
  {Dalcanton}, {Johnson}  \& {PHAT Collaboration}}{{Weisz}
  et~al.}{2012}]{Weisz2012-PHAT}
{Weisz} D.~R.,  {Fouesneau} M.,  {Hogg} D.~W.,  {Rix} H.~W.,  {Dalcanton}
  J.~J.,  {Johnson} L.~C.,   {PHAT Collaboration} 2012, in American
  Astronomical Society Meeting Abstracts \#219. p. 151.06

\bibitem[\protect\citeauthoryear{{Weisz} et~al.,}{{Weisz}
  et~al.}{2015}]{Weisz2015-M31IMF}
{Weisz} D.~R.,  et~al., 2015, \mn@doi [\apj] {10.1088/0004-637X/806/2/198},
  \href {http://adsabs.harvard.edu/abs/2015ApJ...806..198W} {806, 198}

\bibitem[\protect\citeauthoryear{{da Silva}, {Fumagalli}  \& {Krumholz}}{{da
  Silva} et~al.}{2012}]{daSilva2012-SLUGi}
{da Silva} R.~L.,  {Fumagalli} M.,   {Krumholz} M.,  2012, \mn@doi [\apj]
  {10.1088/0004-637X/745/2/145}, \href
  {http://adsabs.harvard.edu/abs/2012ApJ...745..145D} {745, 145}

\bibitem[\protect\citeauthoryear{{da Silva}, {Fumagalli}  \& {Krumholz}}{{da
  Silva} et~al.}{2014}]{daSilva2014-SLUGii}
{da Silva} R.~L.,  {Fumagalli} M.,   {Krumholz} M.~R.,  2014, \mn@doi [\mnras]
  {10.1093/mnras/stu1688}, \href
  {http://adsabs.harvard.edu/abs/2014MNRAS.444.3275D} {444, 3275}

\bibitem[\protect\citeauthoryear{{de Wit}, {Testi}, {Palla}, {Vanzi}  \&
  {Zinnecker}}{{de Wit} et~al.}{2004}]{deWit2004-SolitaryOB}
{de Wit} W.~J.,  {Testi} L.,  {Palla} F.,  {Vanzi} L.,   {Zinnecker} H.,  2004,
  \mn@doi [\aap] {10.1051/0004-6361:20040454}, \href
  {http://adsabs.harvard.edu/abs/2004A%26A...425..937D} {425, 937}

\bibitem[\protect\citeauthoryear{{de Wit}, {Testi}, {Palla}  \&
  {Zinnecker}}{{de Wit} et~al.}{2005}]{deWit2005-SolitaryOB}
{de Wit} W.~J.,  {Testi} L.,  {Palla} F.,   {Zinnecker} H.,  2005, \mn@doi
  [\aap] {10.1051/0004-6361:20042489}, \href
  {http://adsabs.harvard.edu/abs/2005A%26A...437..247D} {437, 247}

\bibitem[\protect\citeauthoryear{{van Dokkum} \& {Conroy}}{{van Dokkum} \&
  {Conroy}}{2010}]{vanDokkum2010-Elipticals}
{van Dokkum} P.~G.,  {Conroy} C.,  2010, \mn@doi [\nat] {10.1038/nature09578},
  \href {http://adsabs.harvard.edu/abs/2010Natur.468..940V} {468, 940}

\makeatother
\end{thebibliography}
